\documentclass[reprint,longbibliography,preprintnumbers,nofootinbib,amsmath,amssymb,aps,onecolumn]{revtex4-2}
\usepackage{dblfloatfix}
\usepackage{graphicx}
\usepackage{dcolumn}
\usepackage{bm}
\usepackage{color}
\usepackage{amssymb}
\usepackage{pifont}
\definecolor{green2}{rgb}{0.0, 0.5, 0.0}
\usepackage[dvipsnames]{xcolor}
\usepackage[colorlinks = true,
            linkcolor = blue,
            urlcolor  = blue,
            citecolor = green,
            anchorcolor = blue]{hyperref}
\usepackage{verbatim}
\makeatletter
\usepackage{multirow, graphicx,amssymb,url,mathrsfs,amsmath}
\usepackage{amsxtra,amstext,latexsym,dsfont,amsfonts}
\usepackage{color,eucal}
\usepackage[dvipsnames]{xcolor}
\usepackage{float}
\usepackage{colortbl}
\usepackage{multirow}
\usepackage{tikz}
\usetikzlibrary{tikzmark}
\usetikzlibrary{arrows}

\newcommand{\Rmnum}[1]{\expandafter\@slowromancap\romannumeral #1@}
\usepackage{tabularx, array}
\newcolumntype{L}[1]{>{\raggedright\arraybackslash}p{#1}}
\newcolumntype{C}[1]{>{\centering\arraybackslash}p{#1}}
\newcolumntype{R}[1]{>{\raggedleft\arraybackslash}p{#1}}
\makeatother

\makeatletter
\newcommand{\myBig}{\bBigg@{1.75}}
\makeatother

\begin{document}
\title{\Large Dynamical evolution of spinodal decomposition\\ in holographic superfluids}
\author{Xin Zhao}
\email{zhaoxin2@stu.kust.edu.cn}
\affiliation{Center for Gravitation and Astrophysics, Kunming University of Science and Technology, Kunming 650500, China}
\author{Zi-Qiang Zhao}
\email{zhaoziqiang@stumail.neu.edu.cn}
\affiliation{Center for Gravitation and Astrophysics, Kunming University of Science and Technology, Kunming 650500, China}
\affiliation{Key Laboratory of Cosmology and Astrophysics (Liaoning) \& College of Sciences,
Northeastern University, Shenyang 110819, China}
\author{Zhang-Yu Nie}
 \email{niezy@kmust.edu.cn}
 \email{Corresponding author.}
\affiliation{Center for Gravitation and Astrophysics, Kunming University of Science and Technology, Kunming 650500, China}
\author{Hua-Bi Zeng}
 \email{hbzeng@yzu.edu.cn}
 \email{Corresponding author.}
\affiliation{Center for Gravitation and Cosmology, College of Physical Science
and Technology, Yangzhou University, Yangzhou 225009, China}
\author{Yu Tian}
 \email{ytian@ucas.ac.cn}
 \email{Corresponding author.}
\affiliation{School of Physical Sciences, University of Chinese Academy of Sciences, Beijing 100049, China \& Institute of Theoretical Physics, Chinese Academy of Sciences, Beijing 100190, China}
\author{Matteo Baggioli}
 \email{b.matteo@sjtu.edu.cn}
 \email{Corresponding author.}
\affiliation{Wilczek Quantum Center, School of Physics and Astronomy, Shanghai Jiao Tong University \& Shanghai Research Center for Quantum Sciences, Shanghai 200240, China}
\date{\today}

\begin{abstract}
We study the nonlinear dynamical evolution of spinodal decomposition in a first-order superfluid phase transition using a simple holographic model in the probe limit. We first confirm the linear stability analysis based on quasinormal modes and verify the existence of a critical length scale related to a gradient instability -- negative speed of sound squared -- of the superfluid sound mode, which is a consequence of a negative thermodynamic charge susceptibility. We present a comparison between our case and the standard Cahn-Hilliard equation for spinodal instability, in which a critical length scale can be also derived based on a diffusive instability. We then perform several numerical tests which include the nonlinear time evolution directly from an unstable state and fast quenches from a stable to an unstable state in the spinodal region. Our numerical results provide a real time description of spinodal decomposition and phase separation in one and two spatial dimensions. We reveal the existence of four different stages in the dynamical evolution, and characterize their main properties. Finally, we investigate the strength of dynamical heterogeneity using the spatial variance of the local chemical potential and we correlate the latter to other features of the dynamical evolution.
\end{abstract}
\maketitle
\tableofcontents
\section{Introduction}\label{Introduction}
\begin{figure}
    \centering
    \includegraphics[width=0.7\linewidth]{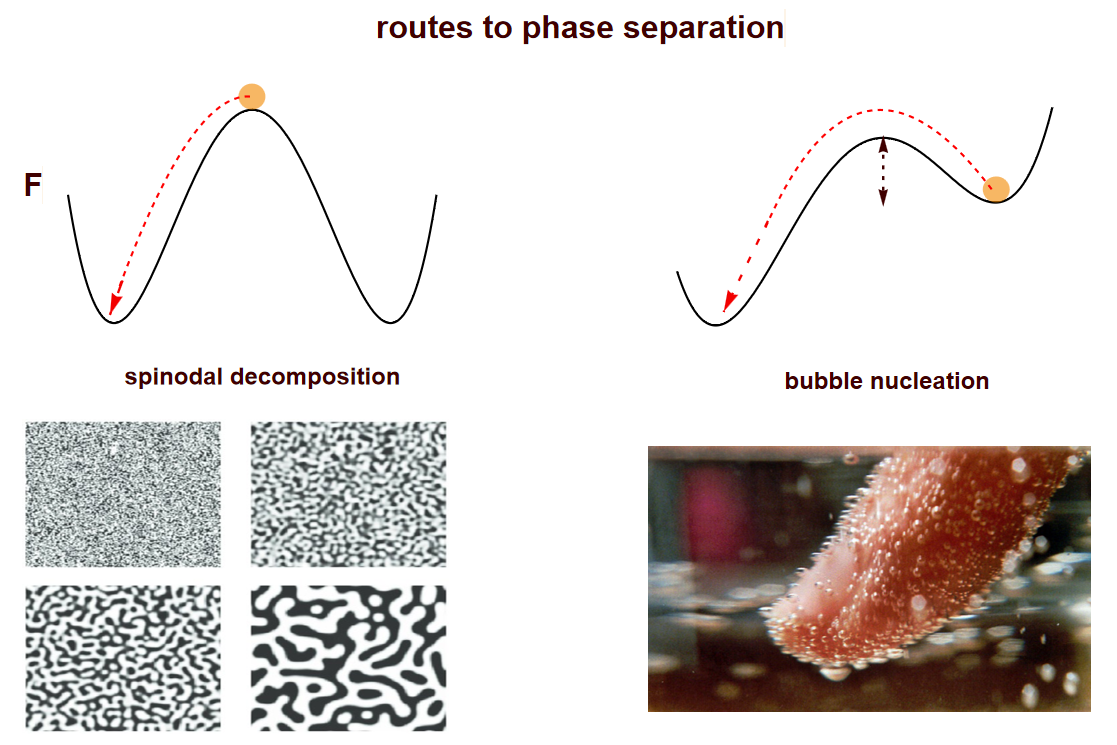}
    \caption{\textbf{The routes towards phase separation in a first-order phase transition.} On the left, the dynamics of spinodal decomposition from a local maximum of the thermodynamic potential $F$ and the corresponding time-dependent dynamics, well described by the Cahn-Hilliard equation. On the right, the dynamics from a metastable local minimum of the potential which evolve via the jump of a thermodynamic barrier indicated with the vertical dashed line. The picture shows an example of bubble nucleation in which bubbles of the thermodynamic favoured phase build inside the thermodynamic disfavoured one, producing phase separation.}
    \label{fig0}
\end{figure}

First (1st) order phase transitions, including for example liquid-gas (evaporation) and solid-liquid (melting) phase transitions, are ubiquitous in nature. Commonly, a 1st order phase transition point is defined as the location at which two phases share the same value of the thermodynamic potential (\textit{e.g.}, free energy). 1st order phase transitions usually occur either from meta-stable states by jumping barriers, or from unstable states following a Cahn-Hilliard linear instability at finite wave-vector. The dynamics related to jumping potential barriers from meta-stable states are known as ``bubble nucleation''. They involve a thermodynamic barrier to phase separation, and they are usually described by Classical Nucleation Theory (CNT) \cite{kalikmanov2012classical,Oxtoby_1992}. On the other hand, the dynamics arising around unstable states, and the consequent time-dependent processes, do not involve any thermodynamic barrier nor nucleation step and they give rise to the so called ``spinodal decomposition'' \cite{Binder_1987}. In both processes, inhomogeneous configurations appear and phase separation is often observed with the production and evolution of bubbles, localized regions of the favoured phase inside the thermodynamically disfavoured phase (see Figure \ref{fig0} for a schematic representation).

The main subject of this work is spinodal decomposition, defined as the phase separation triggered by a linear instability in the spinodal region of a first-order phase transition. On general grounds, spinodal decomposition is well described by Cahn-Hilliard theory \cite{10.1063/1.1744102}. In order to explain the essence of this framework, let us consider the evolution of a system with a conserved charge density $c$ within the mean-field Ginzburg-Landau approximation. At leading order, the free energy can be written as a function of the charge density $c$ (or component concentration) and gradient, as follows~\cite{10.1063/1.1695731}:
\begin{align}
F=\int \left[f(c)+\kappa\,(\nabla c)^2\right]dV, 
\end{align}
where $V$ is the volume of the system and $f$ a scalar thermodynamic function.
Here, $\kappa$ controls the cost in the free energy to produce spatial variations in the concentration $c$, \textit{i.e.}, the energetic cost of inhomogeinities.
With further constraint from the conservation of the total charge, $Q=\int c \,dV$, spinodal decomposition can be described by a generalized diffusion equation:
\begin{equation}
    \frac{\partial c}{\partial t}= M \nabla^2 \mu,  \label{diff}
\end{equation}
where $M$ is the mobility of the system and the chemical potential $\mu$ is defined as $\mu=\partial F/\partial c$. The variation of this free energy functional yields to the Cahn-Hilliard equation for the charge density perturbation which can be expressed in the following form~\cite{10.1063/1.1695731}:
\begin{align}
\frac{\partial c}{\partial t}=M \left(\frac{\partial^2 f}{\partial c^2}\right) \nabla^2 c-2 M \kappa \nabla^4 c.
\end{align}

By going to Fourier space $e^{-i(\omega t-kx)}$, the linearized Cahn-Hilliard equation can be easily solved in terms of the imaginary part of the frequency:
\begin{align}
\mathrm{Im}\left[\omega(k)\right]=-M \chi k^2-2M \kappa k^4, \qquad \text{with}\qquad \chi\equiv \left(\frac{\partial^2 f}{\partial c^2}\right).\label{sol1}
\end{align}
Let us note that the susceptibility $\chi$ is the curvature of the thermodynamic potential $f$. We can see from this formula that, when the susceptibility $\chi$ is negative (\textit{e.g.}, a maximum of the potential $f$), there is a critical wave-vector:
\begin{equation}
    k_c= \sqrt{-\frac{\chi}{2 \kappa}},
\end{equation}
below which the solution in Eq.\eqref{sol1} results in positive values of $\mathrm{Im}\left[\omega(k)\right]$, indicating a linear instability at large length scales (or small $k$). If the system is large enough, the solution in the region with $\chi <0$ will spontaneously undergo the growth of these inhomogeneous perturbations -- spinodal decomposition -- which ultimately results in phase separation. The Cahn-Hilliard theory has been widely used to model spinodal decomposition phenomena from unstable initial states~\cite{PETRISHCHEVA20125481,Petrishcheva2009Exsolution,C9RA01118H} at which (see Fig.\ref{fig0}) the thermodynamic susceptibility is negative.\\

In recent years, holography has been found to be a convenient and effective tool for studying strongly coupled field theories at finite temperatures \cite{zaanen_liu_sun_schalm_2015,hartnoll2018holographic,baggioli2019applied}, especially out of equilibrium \cite{Liu:2018crr}. Through holography, numerical experiments were conducted to study the non-equilibrium physics and phase transition processes in several setups (for a non complete list see Refs.~\cite{Bhaseen:2012gg,Adams:2012pj,Li:2020ayr,Xia:2019eje,Zeng:2019yhi,Yang:2019ibe,PhysRevLett.124.081601,PhysRevLett.129.011602,Li:2019swh,Guo:2018mip,Lan:2023,Yang:2023dvk}). Among the various setups, the holographic superfluid model \cite{Hartnoll:2008vx} has emerged as one of the most attractive platforms to study non-equilibrium physics. To mention some of the main results within this model, non equilibrium processes  were realized in Ref.~\cite{Bhaseen:2012gg}, which also confirmed dynamically the onset of the phase transition indicated by the linear stability analysis. The dynamics of vortexes and turbulence in the holographic superfluid are studied in Ref.~\cite{Adams:2012pj,Lan:2016cgl}, which found that the superfluid kinetic energy spectrum obeys the Kolmogorov -$5/3$ scaling law. The Kibble-Zurek mechanism describing the formation of vortexes is also well tested in the holographic superfluid model~\cite{Sonner:2014tca,Zeng:2019yhi}, and the breaking of its universality in the fast quench processes has interestingly been reported in a recent study~\cite{Zeng:2022hut}.

Back to our main topic, the spinodal decomposition phenomenon has been recently discussed in holographic models~\cite{Janik:2015iry,Janik:2016btb,Janik:2017ykj,Attems:2017ezz,Attems:2019yqn,Bellantuono:2019wbn,Bea:2020ees,Attems:2020qkg,Bea:2021ieq}. 
In Refs.~\cite{Janik:2015iry,Janik:2016btb}, the quasinormal modes (QNMs) of a holographic system with 1st order phase transitions are calculated, providing solid evidence for the linear instability related to spinodal decomposition. The full dynamical evolution in the same model has been then considered in Ref.~\cite{Janik:2017ykj}. The spinodal decomposition triggered by the Gregory-Laflamme instability has been studied in Ref.~\cite{Attems:2017ezz} and the relation with hydrodynamics was discussed. In Ref.~\cite{Attems:2019yqn}, it has been proposed that the time evolution of spinodal decomposition consists of four generic stages: a first, linear stage in which the instability grows exponentially; a second, non-linear stage in which peaks and/or phase domains are formed; a third stage in which these structures merge; and a fourth stage in which the system finally relaxes to a static, phase-separated configuration. Further studies~\cite{Bellantuono:2019wbn,Bea:2020ees,Attems:2020qkg,Bea:2021ieq} also realized spinodal decomposition in various holographic models and studied problems such as the formation and collision of domain walls as well as effects of different values of surface tension. Importantly, the above studies only considered an effective one-dimensional problem, whereas the realization of spinodal decomposition with two effective spacial dimensions was studied only in Ref.~\cite{Bea:2021zol} with a special focus on the emission of gravitational waves. In all these studies, the holographic setup for the 1st order phase transition was an AdS-Einstein-scalar model, which is usually applied to study the 1st order QCD phase transition~\cite{Cai:2022omk}. On the contrary, the bubble nucleation process, involving a finite potential barrier, has been realized in both one and two spacial dimensions in the holographic superfluid model with two competing s-wave orders in Ref.~\cite{Li:2020ayr}, in which the surface tension was calculated and verified. However, the full dynamical processes which accompany spinodal decomposition have not been studied in a holographic superfluid model yet. This will be our main motivation and the main novelty of this work.

Very recently, the linear stability of a simple holographic superfluid model with various kinds of phase transitions including 2nd order, 1st order and 0th order, was studied in detail within the probe limit~\cite{Zhao:2022jvs}. There, linear instabilities at large length scale, with $0<k<k_c$, were discovered in the spinodal region of the 1st order superfluid phase transition. However, differently from the diffusive dynamics in the Cahn-Hilliard theory presented above, the instabilities in this superfluid model turned out to have a different origin. Indeed, it is well known that a superfluid presents dynamics which are richer than the conservation of a single charge, due to the emergence of Goldstone degrees of freedom arising because of the spontaneous symmetry breaking of the U(1) symmetry. In particular, its hydrodynamic description cannot be correctly formulated ignoring this fact. In a relativistic superfluid, the U(1) charge does not diffuse anymore, as envisaged in Eq.\eqref{diff}, but it rather couples to the Goldstone mode (\textit{i.e.}, the fluctuations of the phase of the superfluid order parameter) creating a propagating wave-like excitation known as 2nd sound \cite{10.1063/1.3248499}. In the broken phase, the dispersion of this mode at low wave-vector is given by
\begin{align}\label{SoundMomega}
\omega_{\pm}(k) = \pm c_s k - i\Gamma_{s} k^2 +\dots 
\end{align}
and has been studied in detail in the holographic superfluid model \cite{Arean:2021tks,Amado:2009ts,Herzog:2011ec,Herzog:2009md,Ammon:2021pyz}.

Hence, in a superfluid, the instability which leads to spinodal decomposition arises because the second sound speed square $c_s^2$ becomes negative. When that happens, the sound dispersion takes the following form
\begin{align}\label{SoundMomega2}
\omega_{\pm}(k) =i \left( \pm |c_s| k -\Gamma_{s} k^2\right)+\dots
\end{align}
which implies the presence of a critical wave-vector:
\begin{equation}
    k_{c}=|c_{s}|/\Gamma_{s} 
\end{equation}
below which:
\begin{equation}
    \mathrm{Im}\left[\omega_{+}(k)\right]>0 \qquad \text{for}\qquad k<k_c.
\end{equation}

A negative $c_s^2$ is a consequence of a negative thermodynamic susceptibility $\chi<0$. Indeed, from superfluid hydrodynamics it is well known that $c_s^2\propto \chi^{-1}$ \cite{Herzog:2011ec,Arean:2021tks}. Although this instability arises from the growth of the superfluid sound mode, rather than a diffusive mode as considered in Eq.\eqref{diff}, the similarities with Cahn-Hilliard theory are striking. First, both cases concern an inhomogeneous instability which develops only below a critical wave-vector. Moreover, in both cases the instability is in a sense thermodynamic since it corresponds to a thermodynamic susceptibility becoming negative.

In summary, in the first-order holographic superfluid model, we do expect an instability with $\mathrm{Im}(\omega)>0$ for $0<k<k_{c}$. This behavior is reminiscent of the well-known Gregory-Laflamme type instability of black strings~\cite{Gregory:1993vy}, and it is qualitatively similar to the sound mode instabilities observed in the previous holographic studies which do not involve a superfluid system~\cite{Janik:2016btb,Attems:2017ezz,Janik:2017ykj,Attems:2019yqn,Bellantuono:2019wbn,Bea:2020ees,Attems:2020qkg}. Based on these progresses, our aim is to extend the previous analyses by considering the real-time dynamics of spinodal decomposition in  the holographic superfluid model of~\cite{Zhao:2022jvs}. For simplicity, we will work in the probe limit, where there are less numerical challenges and the formation and evolution of 2-dimensional round bubbles can be observed and tracked in an easier way.\\

The structure of the paper is as follows: in Section~\ref{sec1}, we introduce the holographic model and briefly review the static solutions and the QNMs of the model. In Section~\ref{sec2}, we perform the ``numerical experiments''. We test the linear stability of homogeneous initial states in the spinodal region, and we confirm the theoretical expression for the critical length scale. Moreover, we perform quenching experiments to show the full dynamical processes of spinodal decomposition in the dual (2+1)-dimensional field theory. Finally, we provide some comments on the question of whether the final state is uniform or not. In Section~\ref{sect:conclusion}, we summarize and present an outlook of our results.
\section{Holographic setup and linear stability}  \label{sec1}
\subsection{Holographic setup and the 1st order phase transition}\label{subsec1}
We consider the holographic s-wave superfluid model with nonlinear self-interaction terms in (3+1) dimensional asymptotic AdS spacetime~\cite{Zhao:2022jvs}, which is a simple extension of the HHH model \cite{Hartnoll:2008vx}. The bulk action of this model is given by
\begin{align}
     S=&\,S_{M}+S_{G},\label{Lagall}\\
     S_G=&\,\frac{1}{2\kappa_g ^2}\int d^{4}x\sqrt{-g}(R-2\Lambda),\label{Lagg}\\
     S_M=&\int d^{4}x\sqrt{-g}\Big(-\frac{1}{4}F_{\mu\nu}F^{\mu\nu}
	-D_{\mu}\Psi^{\ast}D^{\mu}\Psi-m^{2}\Psi^{\ast}\Psi-\lambda(\Psi^{\ast}\Psi)^{2}-\tau(\Psi^{\ast}\Psi)^{3}\Big).\label{Lagm}
\end{align}

Here, $F_{\mu\nu}=\nabla_{\mu}A_{\nu}-\nabla_{\nu}A_{\mu}$ is the field strength of the U(1) gauge field, and $D_{\mu}\Psi=\nabla_{\mu}\Psi-i q A_\mu\Psi$ the standard covariant derivative. $\Lambda=-3/L^2$ is the negative cosmological constant, where $L$ is the AdS radius. $\Psi$ is a complex scalar field charged under the U(1) gauge symmetry. The additional two terms appearing in the matter Lagrangian, Eq.\eqref{Lagm}, $\lambda(\Psi^{\ast}\Psi)^{2}$ and $\tau(\Psi^{\ast}\Psi)^{3}$, introduce nonlinear self-interactions for the bulk scalar field $\Psi$.

The influence of the nonlinear term $\lambda (\Psi^{\ast}\Psi)^{2}$ has been investigated analytically in Ref.~\cite{Herzog:2010vz} and  numerically in a holographic s+p model in asymptotic AdS$_{4}$ spacetime in Ref.~\cite{Zhang:2021vwp}. In Ref.~\cite{Zhao:2022jvs}, both the two nonlinear terms $\lambda (\Psi^{\ast}\Psi)^{2}$ and $\tau (\Psi^{\ast}\Psi)^{3}$ were considered and the corresponding QNMs spectrum analyzed. Thanks to these additional terms, the phase diagram of the model becomes very rich and it includes 2nd, 1st as well as 0th order phase transitions. 

In the rest of this paper, for simplicity, we fix $m^2=-2$, $q=1$, $z_{h}=1$, $L=1$. The values of the two nonlinear parameters are set to $\lambda=-2$, $\tau=0.8$ and $\lambda=-0.8$, $\tau=0.28$, in order to focus on two typical 1st order phase transitions. In the rest of this section, we briefly review previous results concerning the thermodynamics and the linear stability of this model for the choice of parameters $\lambda=-2$, $\tau=0.8$, which were already presented in Ref.~\cite{Zhao:2022jvs}.

In the probe limit, the background geometry is given by a (3+1) dimensional black brane
\begin{align}
    &\ ds^{2}=\dfrac{L^2}{z^2}\left(-f(z)dt^{2}+\frac{1}{f(z)}dz^{2}+dx^{2}+dy^{2}\right),\label{metric1}\\
    &\ f(z)=1-\left(z/z_h\right)^3.
\end{align}

In these notations, $z$ is the radial coordinate in the bulk, with $z=0$ the location of the AdS boundary and $z=z_{h}$ the horizon. The Hawking temperature of this black brane solution is given by
\begin{align}
    T= \frac{3}{4\pi z_h}.
\end{align}
For homogeneous static solutions, we consider the following ansatz for the matter fields:
\begin{align}\label{ansatz1}
    \Psi=z \psi(z)/L, \quad A_{\mu}dx^{\mu}=A_{t}(z)dt=\phi(z)dt,
\end{align}
which results in the following equations of motion
\begin{align}
    \psi''+\dfrac{f'}{f}\psi'-\dfrac{z}{f}\psi+\dfrac{\phi^{2}}{f^{2}}\psi-\dfrac{2\lambda}{f} \psi^{3}-\dfrac{3\tau z^{2}}{f}\psi^{5}=0,\label{eqpsi}\\
    \phi''-\dfrac{2\psi^{2}}{f}\phi=0,\label{eqphi}
\end{align}
where the prime denotes the derivative with respect to $z$.

To solve these coupled equations, we need to specify the boundary conditions. The expansions for the bulk fields near the horizon $z\rightarrow z_{h}$ are given by
\begin{align}
    &\phi(z)=\phi_{1}(z-z_{h})+\mathcal{O}((z-z_{h})^{2}),\\
    &\psi(z)=\psi_{0}+\mathcal{O}(z-z_{h}).
\end{align}
On the contrary, near the AdS boundary $z\rightarrow 0$ the bulk fields are of the form
\begin{align}
    &\phi(z)=\mu-z \rho+...,\qquad \\
    &\psi(z)=\psi^{(1)}+z \psi^{(2)}+...,
\end{align}
where $\mu$ and $\rho$ are the chemical potential and charge density of the boundary system, respectively. We follow the standard quantization scheme, in which $\psi^{(1)}$ is regarded as the source term for the boundary operator while $\psi^{(2)}$ is regarded as the vacuum expectation value. We set the source free condition $\psi^{(1)}=0$ to obtain the solutions corresponding to the spontaneous symmetry breaking of the boundary global $U(1)$ symmetry.

Using the AdS/CFT dictionary, the free energy coincides with the Euclidean on-shell action with proper boundary terms, 
\begin{align}
	F=\frac{V_2 L^{2}}{T}\left(\frac{\mu\rho}{2}+\int_{0}^{z_h}\left(\frac{q^2 \phi^2 \psi^2}{f}-\lambda\psi^4-2z^2\tau\psi^6\right)dz\right).
\end{align}
This thermodynamic potential involves only the matter contribution because of the probe limit and it is valid only for homogeneous solutions.

\begin{figure}
      \includegraphics[width=0.30\columnwidth]{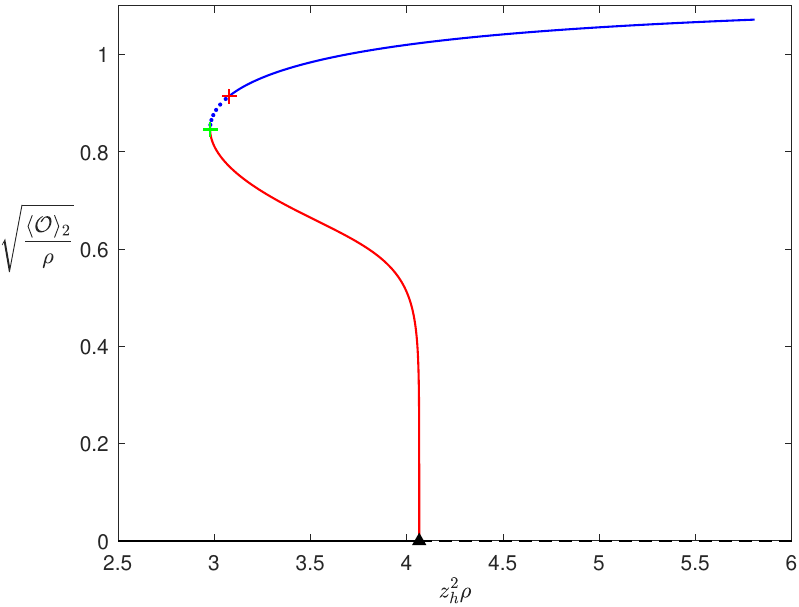}
      \includegraphics[width=0.30\columnwidth]{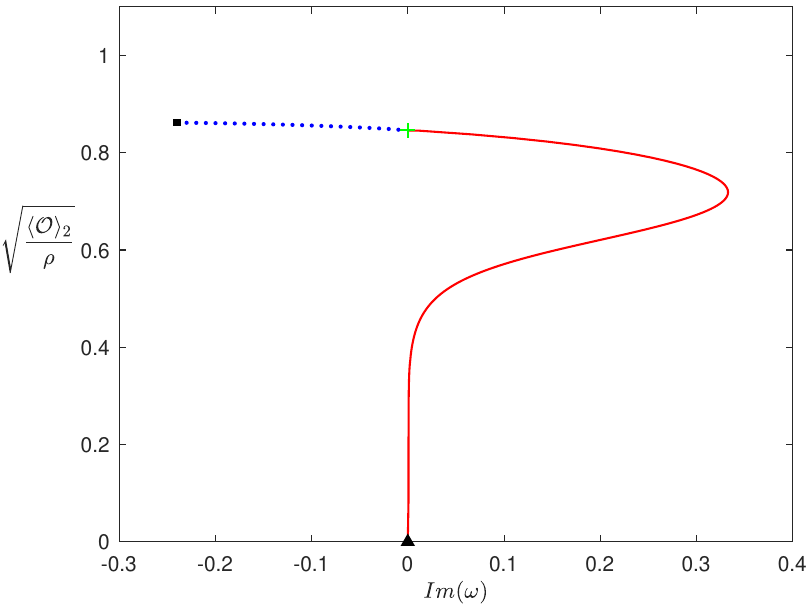}
      \includegraphics[width=0.30\columnwidth]{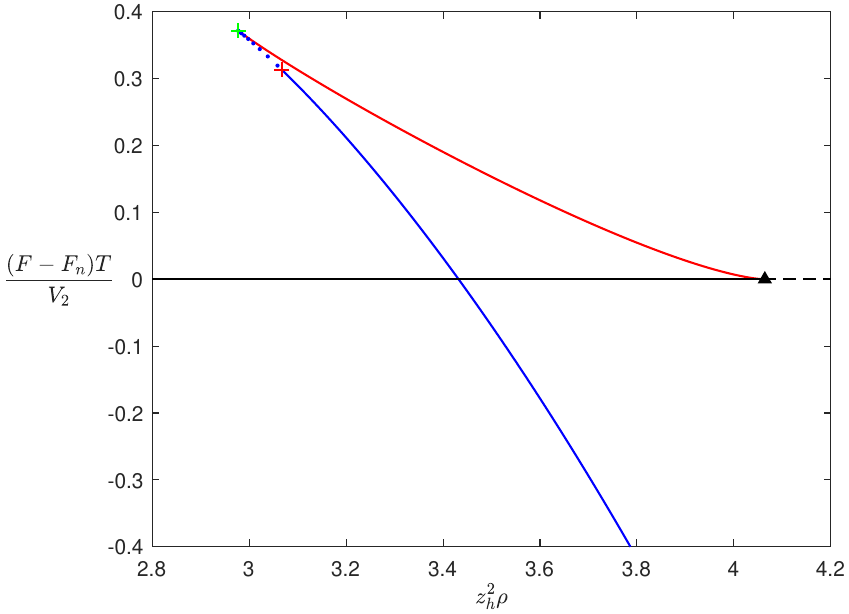}
\caption{\textbf{Static homogeneous solutions with $\lambda=-2$ and $\tau=0.8$}.
\textbf{Left: }The condensate as a function of the charge density $\rho$.
\textbf{Center: }The relation between the imaginary part $\mathrm{Im}(\omega)$ of the amplitude mode and the condensate.
\textbf{Right: }The free energy difference between the normal solution and the superfluid solution. The black triangle denotes the ``critical point'' at which the broken phase connects to the normal one. The green cross denotes the turning point of the condensate curve in the canonical ensemble, while the red cross denotes the turning point of the condensate curve in the grand canonical ensemble. The black square denotes the location where the amplitude mode collides with another pure imaginary mode (see \cite{Zhao:2022jvs} for details). The solid red line is the unstable branch of solutions, while the solid blue is either stable or meta-stable. Finally, the dotted blue part is stable under homogeneous perturbations but unstable under inhomogeneous ones.}\label{1st_static}
\end{figure}

Next, we show the results for the choice of parameters $\lambda=-2$, $\tau=0.8$, whose linear dynamical stability has already been presented in Ref.~\cite{Zhao:2022jvs}. The superfluid condensate, the imaginary part $\mathrm{Im}(\omega)$ of the amplitude mode as well as the free energy with respect to the charge density $\rho$ are plotted in the left, central, and right panels of Figure~\ref{1st_static}, respectively. The other case, $\lambda=-0.8$, $\tau=0.28$, will be consider later on.

The condensate at the critical point first grows leftwards up to the turning point (green cross) and then changes direction, rather than directly growing rightwards at the critical point as in second order phase transitions. Notice that the imaginary part of the frequency of the amplitude mode is positive along the red branch, implying the linear instability of those solutions. The free energy curve forms a ``swallow" tail shape and has discontinuous first derivatives at the phase transition point, which is defined as the intersection point of the black and blue curves in the free energy. All these features clearly demonstrate the presence of a first order phase transition in this system.

\subsection{Quasinormal modes (QNMs) and linear stability}\label{subsec2}
Taking advantage of the gauge/gravity correspondence, the linear stability of the dual quantum system under infinitesimal perturbations can be studied by analyzing the QNMs of the classical gravity dual. In this subsection, we review the results for the 1st order phase transition with $\lambda=-2$ and $\tau=0.8$ obtained in Ref.~\cite{Zhao:2022jvs}.
In the QNM analysis, if the imaginary part of the frequency $\omega$ of one mode is positive, this mode will grow exponentially in time $\sim \exp(|\mathrm{Im}(\omega)|t)$ destabilizing the original background solution. Therefore, only when the imaginary part of $\omega$ of all the modes is negative, the background solution is linearly stable. In other words, all the possible linear perturbations will not grow up in time but eventually relax back to the background solution. In general, the complex frequency $\omega$ also depends on the real wave vector $k$. Homogeneous perturbations corresponds to $k=0$ and constitute only a sub-set of all possible kind of linear perturbations. In order to test completely the linear stability of the system around a certain background solution, the full spectrum of $\omega(k)$ has to be considered.

The results for the imaginary frequency of the amplitude mode with $k=0$ are shown in the central panel of Figure~\ref{1st_static}. They show that under homogeneous perturbations, the superfluid solution is unstable below the turning point of the condensate curve (red branch in the left panel of Figure~\ref{1st_static}), and stable above (blue branch in the left panel of Figure~\ref{1st_static}). This fits perfectly with the analysis from the free energy landscape, right panel of Figure~\ref{1st_static}.

\begin{figure}
	\includegraphics[width=0.33\columnwidth]{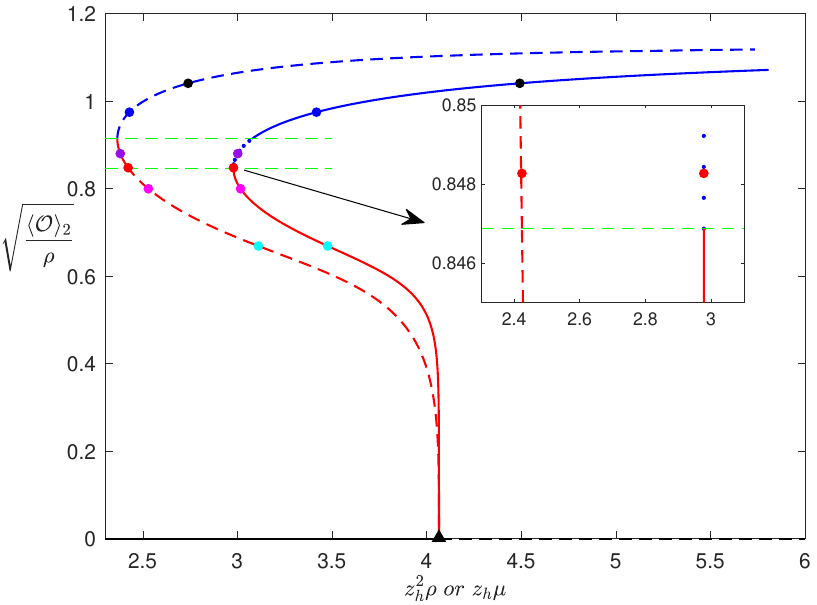}
        \includegraphics[width=0.31\columnwidth]{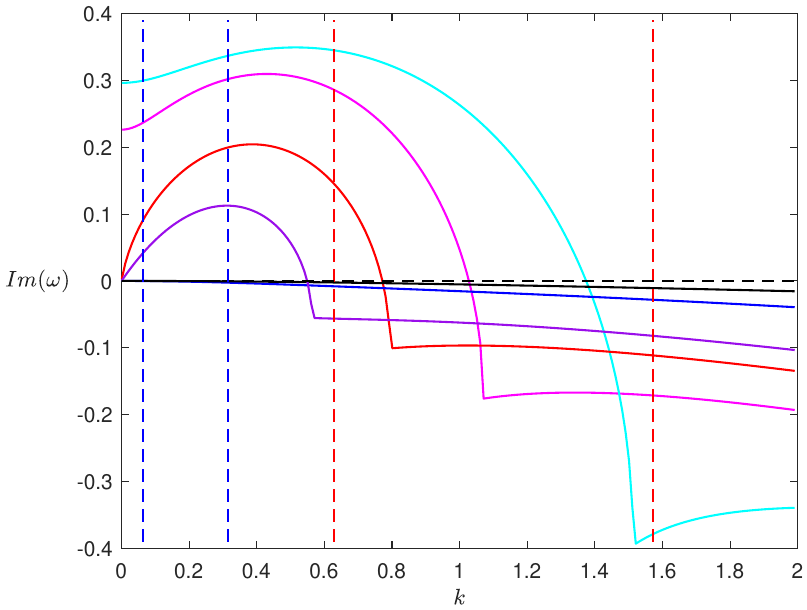}
	\includegraphics[width=0.32\columnwidth]{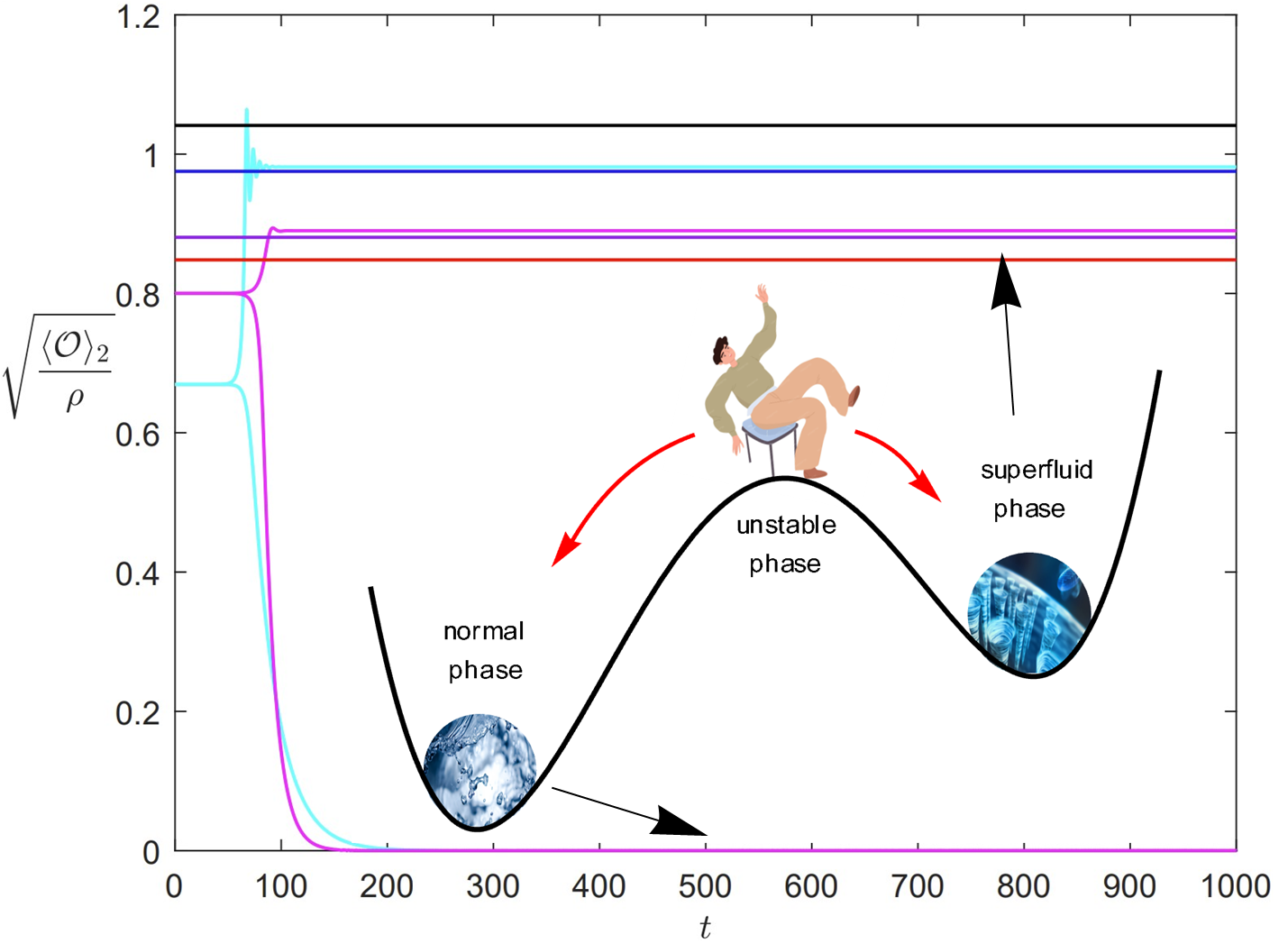}
	\caption{\textbf{Left: }The condensate as a function of the charge density $\rho$ (dashed line) or chemical potential $\mu$ (solid line). The two lines correspond respectively to the canonical ensemble and the grand canonical ensemble. The black triangle represents the critical point. The two dashed green horizontal lines denote the turning points in the two different ensembles.
      \textbf{Center: }The dependence of $\mathrm{Im}(\omega)$ on $k$ for the most unstable mode around different solutions. The dashed black line is the horizontal axes with $\mathrm{Im}(\omega)=0$. The two dashed red vertical lines indicate the two spatial values of the wave-vector which are related to the two spacial scales $l_x$=10 and 4, respectively, used in the dynamical tests. The two dashed blue vertical lines indicate $k_{1}=2\pi/100$ and $k_{5}=5*2\pi/100$ for the system with $l_x=100$, respectively.
		 \textbf{Right: }The time dependent condensate of the uniform initial solution in the real-time dynamical homogeneous evolution. The uniform initial solution is marked with the same color in the left panel. The inset is a cartoon to illustrate the two different evolutions for the magenta and cyan curves.}\label{1st_dynamic_k0}
\end{figure}

The results for the QNMs at finite $k$ are more interesting. We select six different solutions marked by different colored points in the left panel of Figure~\ref{1st_dynamic_k0}. In the corresponding central panel, we show the imaginary part of the lowest mode as a function of the wave-vector $k$ for each of these six solutions. We observe that the solutions marked with black and blue points in the left panel are stable for any value of the wave-vector $k$, since the imaginary part of this mode (and indeed of all other modes) is negative. On the contrary, the solutions marked by cyan, magenta, red, and violet colors are unstable. More precisely, for the red and purple points, the corresponding solution is stable under homogeneous $k=0$ perturbations but unstable under perturbations with finite wave-vector. Finally, the solutions marked with magenta and cyan colors in the left panel of Figure~\ref{1st_dynamic_k0} are unstable under both $k=0$ and $k \neq 0$ perturbations. As already anticipated in \cite{Zhao:2022jvs}, this is closely related to the difference between the canonical and grand canonical ensembles. To better illustrate this point, we plot the condensate curves in the two ensembles with solid and dashed curves respectively in the left panel of Figure~\ref{1st_dynamic_k0}, where we use dashed green lines to denote the two turning points for the two different ensembles. Importantly, these horizontal dashed lines correspond the onset of linear instability at $k=0$ and $k \neq 0$ respectively. The canonical ensemble agrees well with the linear stability analysis at $k=0$ but not with the stability analysis under inhomogeneous perturbations, which is on the contrary captured by the grand canonical ensemble. We can see this explicitly by looking at the purple solution between the two different curves. In the canonical ensemble the derivative of the condensate is positive, indicating thermodynamic stability. On the contrary, in the grand-canonical ensemble the derivative is negative hinting towards a thermodynamic instability. At the same time, the QNM analysis reveals that such a solution is stable at $k=0$, as one could derive from the canonical ensemble, but unstable at $k\neq 0$, as hinted by the grand-canonical ensemble. For more details on this point see \cite{Zhao:2022jvs}.

It is worth noticing that the imaginary part of the unstable mode crosses the horizontal axes at a critical value $k=k_c$ and the mode becomes stable again at $k>k_c$. Using the numerical data, we derive a critical length scale $l_c=2\pi/k_c$. This scale sets the onset of the instability driven by the inhomogeneous perturbations. Importantly, these unstable modes are sound modes described by Eq.\eqref{SoundMomega}, which share the same qualitative feature with the Gregory-Laflamme instability of black rings.
The study of the QNMs also tell us that, at large enough length scales, the solutions between the two turning points (between the two green dashed lines in Figure~\ref{1st_dynamic_k0}) will be destabilized by the growth of unstable inhomogeneous modes -- spinodal decomposition. However, the QNM analysis does not tell us anything about the non-linear time dependent dynamics nor the endpoint of the instability. This will be the topic of next section.
\section{Numerical experiments and nonlinear dynamical evolution}\label{sec2}
In this section, we simulate the full dynamical processes. Let us start with a brief overview of the setup used and the numerical methods.
\subsection{Holographic setup for the dynamical tests}\label{subsec3}
To simplify the numerical calculation, it is convenient to use in-going Eddington coordinates
\begin{align}
	ds^{2}=\dfrac{L^2}{z^2}\left(-f(z)dt^{2}-2dtdz+dx^{2}+dy^{2}\right)\label{metric2}.
\end{align}

In this choice of coordinates, the component $A_{z}$ of the U(1) gauge field is nonzero. Nevertheless, we can set $A_{z}=0$ using the gauge symmetry, at the cost of having a complex valued scalar field $\Psi$. For general time dependent solutions, it is consistent to use the following ansatz
\begin{align}\label{ansatz2}
	\Psi=z \psi(z,\vec{x},t)/L~, \quad A_{\mu}dx^{\mu}=A_{t}(z,\vec{x},t)dt+A_{\vec{x}}(z,\vec{x},t)d\vec{x}=\phi(z,\vec{x},t)dt+A_{\vec{x}}(z,\vec{x},t)dx~,
\end{align}
where $\vec{x}$ includes space directions $x$ and $y$, and $\psi(z,\vec{x},t)$ is a complex valued function. 

From the lagrangian, Eq.\eqref{Lagm}, we get the time dependent equations of motion
\begin{align}
	2\partial_z\partial_t\psi-f\partial_z\partial_z\psi-f'\partial_z\psi-i\psi\partial_z\phi-2i\phi\partial_z\psi+z \psi+2\lambda\psi^{2}\psi^{*}+3\tau z^{2}\psi^{3}\psi^{*2}-\partial_x \partial_x \psi-\partial_y \partial_y \psi \quad\quad\nonumber\\
	+i\psi\partial_x A_x+i\psi\partial_y A_y+2iA_x\partial_x \psi+2iA_y\partial_y \psi+A_{x}^{2}\psi+A_{y}^{2}\psi=0~,\label{equation1}\\
	\partial_z\partial_t\phi+2\phi\psi\psi^{*}+i\psi^{*}\partial_t\psi-i\psi\partial_t\psi^{*}+if\psi\partial_z\psi^{*}-if\psi^{*}\partial_z\psi-\partial_x \partial_x \phi-\partial_y \partial_y \phi \quad\quad\nonumber\\
	+\partial_t\partial_x A_x+\partial_t\partial_y A_y-f\partial_z\partial_x A_x-f\partial_z\partial_y A_y=0~,\label{equation2}\\
	2\partial_z\partial_tA_x-f\partial_z\partial_zA_x-\partial_z\partial_x\phi-f'\partial_zA_x+2A_x\psi\psi^{*}-\partial_y\partial_yA_x \quad\quad\nonumber\\
	+\partial_x\partial_yA_y-i\psi\partial_x\psi^{*}+i\psi^{*}\partial_x\psi=0~,\label{equation3}\\
	2\partial_z\partial_tA_y-f\partial_z\partial_zA_y-\partial_z\partial_y\phi-f'\partial_zA_y+2A_y\psi\psi^{*}-\partial_x\partial_xA_y \quad\quad\nonumber\\
	+\partial_x\partial_yA_x-i\psi\partial_y\psi^{*}+i\psi^{*}\partial_y\psi=0~,\label{equation4}\\
	\partial_z\partial_z\phi-\partial_z\partial_x A_x-\partial_z\partial_y A_y+i\psi\partial_z\psi^{*}-i\psi^{*}\partial_z\psi=0~.\label{equation5}
\end{align}

In order to solve these equations, we consider a finite volume boundary system in a square box with periodic boundary conditions in the $x$ and $y$ directions. The dependence on these two directions is expanded in Fourier spectrum with $\Delta x=0.25$. The $z$ dependence of the functions is expanded using a Chebyshev spectrum with $21$ grid points. Finally, we apply the fourth-order Runge-Kutta (RK4) method to promote the time evolution with a time step $\Delta t=0.05$.

Besides the periodic boundary conditions in the $x$ and $y$ directions, we also need to specify the boundary conditions along the $z$ direction. Eq.\eqref{equation2} is a constraint equation; we take its value on the conformal boundary
\begin{align}
	&\partial_t\rho=-\partial_x(\partial_zA_x+\partial_x \phi)-\partial_y(\partial_zA_y+\partial_y \phi)~
\end{align}
as the boundary condition for the bulk field $\phi$ along the $z$ direction. This condition is equivalent to the conservation of U(1) charge on the boundary system. Finally, we impose Dirichlet boundary conditions for $\psi$ and $A_\mu$ at $z=0$. After specifying the equations of motion and boundary conditions, we are ready to run the numerical experiment and observe the full time dependent evolution of this holographic superfluid system along the spinodal decomposition.
\subsection{Testing linear stability with dynamical processes}
In the first stage of the numerical experiments, we test the linear stability results obtained from QNMs, by setting an initial state and see whether it stays still or go away to another configuration in a very long time. From this setup, the critical scale defined in the previous section from the wave-vector at which the mode becomes unstable is also confirmed.
\subsubsection{Homogeneous perturbations}\label{subsec4}
We start testing the stability of initial homogeneous states under homogeneous perturbations, which should verify the results of the QNMs with $k=0$. In this group of numerical experiments, we set the time dependent fields as well as perturbations to only depend on $t$ and $z$. We set the initial condition to be a static solution and let it evolve following the non-linear dynamics.

We choose six typical static homogeneous solutions marked by the different colored points in the left panel of Figure~\ref{1st_dynamic_k0} as the initial states of the dynamic evolution. We show the value of condensate as a function of time in the right panel of Figure~\ref{1st_dynamic_k0}. For convenience, we also show the detailed information for the six typical solutions used as initial states in the ``numerical experiments``, as well as the turning point in the canonical ensemble and the grand canonical ensemble marked by the green cross and red cross and the critical point marked by the black triangle in Table~\ref{specific_data}.
\begin{table}
	\begin{tabular}{| C{1.5cm} | C{1.5cm} | C{1.5cm} | C{2.0cm}  | C{1.5cm} | C{1.5cm} | C{2cm}  | C{1.5cm} | C{1.5cm}  | C{2.0cm}  |}
		\hline
		  &    black    &    blue    &    red cross     &    violet    &    red    &    green cross    &    magenta    &    cyan    &    black triangle    \\
		\hline
		$\rho$    &    4.4909    &    3.4164    &   3.0763    &    3.0000    &    2.9775    &    2.9774    &    3.0152    &    3.4754    &    4.0637    \\
		\hline
		$\mu$    &    2.7379    &    2.4276    &   2.3638    &    2.3792    &    2.4207    &    2.4231    &    2.5278    &    3.1092    &    4.0637    \\
		\hline
		$\sqrt{\langle\mathcal{O}\rangle_{2} / \rho}$    &    1.0413    &    0.9753    &   0.9146    &    0.8806    &    0.8483    &    0.8469    &    0.8002    &    0.6694    &    0    \\
		\hline
	\end{tabular}
	\caption{\textbf{The different initial states considered in the ``numerical experiments'' and the turning points.} Specific values of $\rho$, $\mu$ and $\sqrt{\langle\mathcal{O}\rangle_{2} / \rho}$ for the different colored points in Figures \ref{1st_static} and Figures \ref{1st_dynamic_k0}. The black triangle is used to mark the ``critical point'' in both the two figures. The red and green crosses can be found in Figure \ref{1st_static}, and other points can be found in the left panel of Figure \ref{1st_dynamic_k0}.}\label{specific_data}
\end{table}

We can see from the right panel of Figure~\ref{1st_dynamic_k0} that the red, violet, blue, and black curves are horizontal lines with no time dependence. This implies that the four related solutions marked by the red, violet, blue, and black points in the left panel of Figure~\ref{1st_dynamic_k0} are therefore stable upon linear perturbations. For the solutions marked by the magenta and cyan points, the dynamical evolution shows that they are unstable under very small homogeneous perturbations. Indeed, after a finite amount of time, the two solutions run away from the initial state. It is interesting to notice that there are two different paths for each initial state, which also end in different final states. This is because the initial unstable states are saddle points in the potential landscape, where on one side is the way ``down the hill'' towards the (meta-)stable normal phase solution, while on the other side is the way ``down the hill'' towards the (meta-)stable superfluid solution on the upper branch of the condensate curve (see inset in the right panel of Figure~\ref{1st_dynamic_k0} for a cartoon). Indeed, at late time, one of the two branches reaches a final state with zero condensate, while the other a final state with a constant and finite condensate, different from that of the initial state. We have further confirmed that the difference between the two paths, and the corresponding final states, is related to the sign of the perturbations of the scalar field around the initial states. In addition, we observe that when the final state is a superfluid state with finite condensate, the cyan and magenta curves show an oscillating behavior while approaching the equilibrium final state. This phenomenon is consistent with the results of QNMs of the final state. In particular, it is simply the sign that the late time relaxational dynamics are dominated by the lowest QNM -- the superfluid sound -- (see \cite{PhysRevLett.110.015301} for similar findings). These oscillations are not present when the time evolution ends in the normal phase, since the latter does not contain any sound mode in the spectrum (in the probe limit).

The above numerical experiments show that the solutions marked by the red, violet, blue, and black points are stable under homogeneous perturbations, while the the solutions marked by the magenta and cyan points are unstable, which is consistent with the results of the $k=0$ quasinormal modes, as well as the potential landscape analysis in the canonical ensemble.
\subsubsection{Inhomogeneous case}\label{subsec5}
To continue, we now focus on the numerical experiments regarding the stability of the initial homogeneous states under inhomogeneous perturbations, which will be compared with the results from the QNMs at finite $k$. In this set of numerical experiments, for simplicity, we still turn off the $y$ dependence, which means the dynamical fields as well as perturbations depends only on $t,z,x$.

\begin{figure}
	\includegraphics[width=0.41\columnwidth]{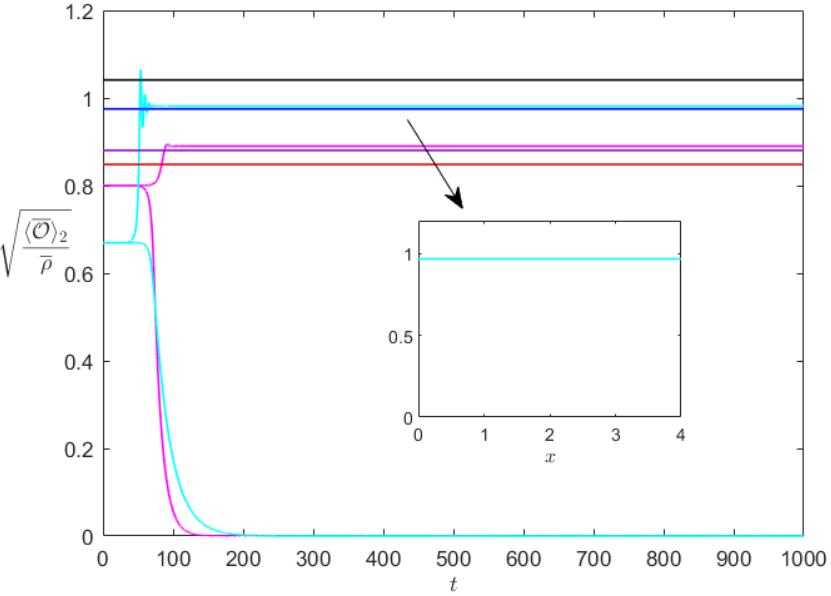}
	\includegraphics[width=0.4\columnwidth]{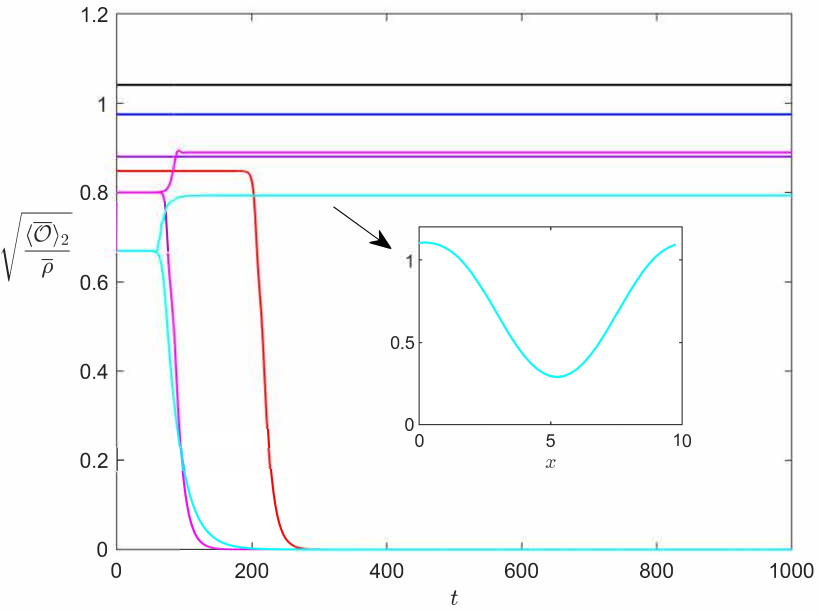}
	\caption{The time dependent value of the average condensate from different initial solutions when the inhomogeneous states are involved in the time evolution. The size of the system along the $x$ direction is set to $l_x=4$ for the \textbf{left panel} and $l_x=10$ for the \textbf{right panel}. 
 The lines with different colors represent the time evolution from different initial solutions. The colors are the same as in the left plot of Figure~\ref{1st_dynamic_k0}. The condensate values are the average in the $x$ direction. The insets show the configuration of the local condensate in the final state for the cyan solution.}\label{1st_dynamic_kneq0}
\end{figure}

As before, we take the six typical static homogeneous solutions marked by different colored points in the left panel of Figure~\ref{1st_dynamic_k0} as the initial states of the dynamic evolution (see Table \ref{specific_data} for details), and show the average value of condensate as a function of time in Figure~\ref{1st_dynamic_kneq0}. The left and right panels correspond respectively to $l_x=4$ and $l_x=10$, where $l_x$ is the size of the system along the $x$ direction. 

We can see that the left panel of Figure~\ref{1st_dynamic_kneq0} is almost the same as the right panel of Figure~\ref{1st_dynamic_k0}. This is because the length scale of the system $l_x=4$ is very small and indeed below the onset of instability shown in the central panel of Fig.\ref{1st_dynamic_k0}. In order to make this clearer, we use the relation between the wave vector and the wave length, $k=2\pi/l$, to calculate the two values of $k_x$ related to $l_x=4$ and $l_x=10$. Here, $k_x$ is the minimum wave-vector allowed in a box of size $l_x$. These two values are marked with dashed red vertical lines in the central panel of Figure~\ref{1st_dynamic_k0}. The dashed right line corresponds to $l_x=4$ and the left one to $l_x=10$. We clearly see that the value of $k_x$ corresponding to $l_x=4$ is larger than all the critical wave-vectors. In other words, since only modes with $k>k_x$ are allowed in this system, no unstable modes can appear. As a consequence, the solutions marked by the red and violet points are stable. However, the finite size does not rule out the homogeneous mode with $k=0$ since all allowed wave-vectors are given by $2 n\pi/l_x, (n=0,1,2...)$. Hence, the solutions marked by the cyan and magenta points are still unstable. 

Now, let us look at the right panel of Figure~\ref{1st_dynamic_kneq0}, where the length scale $l_x$ is larger and part of the finite $k$ unstable modes for the solutions marked by the cyan, magenta, and red points are allowed. The purple solution is stable because the system is too small to contain the corresponding unstable modes (see central panel of Figure~\ref{1st_dynamic_k0}). As shown by the red curve, the solution marked by the red point is not stable and it decays into the normal phase. The magenta curve is the same as the left panel of Figure~\ref{1st_dynamic_kneq0}, while one of the cyan curves shows a new process approaching a final state with a smaller average value of condensate. We expect that this new final state is inhomogeneous. In order to confirm this, we plot the local value of condensate as a function of the $x$ coordinate in the inset of the right panel in Figure~\ref{1st_dynamic_kneq0}, where we see a clear $x$ dependence in this final state. The inset of the left panel in Figure~\ref{1st_dynamic_kneq0} shows the case of the final state of the upper cyan curve in the left plot of Figure~\ref{1st_dynamic_kneq0} as a comparison.

The 1-dimensional non uniform final state shown in the inset in the right panel of Figure~\ref{1st_dynamic_kneq0} is not a typical 1-dimensional bubble state. More precisely, the region inside of the bubble still presents a finite condensate and therefore is not in the normal phase. In order to better explore the formation and evolution of bubbles, in the next subsection, we try different initial states and we study the dynamical quenching processes in 2-dimensional boundary space. Before that, let us discuss in more detail the critical length and perform more dynamical tests to confirm the value of this critical length $l_c$ calculated from the critical wave vector $k_c$ of the QNMs.

\begin{figure}
	\includegraphics[width=0.4\columnwidth]{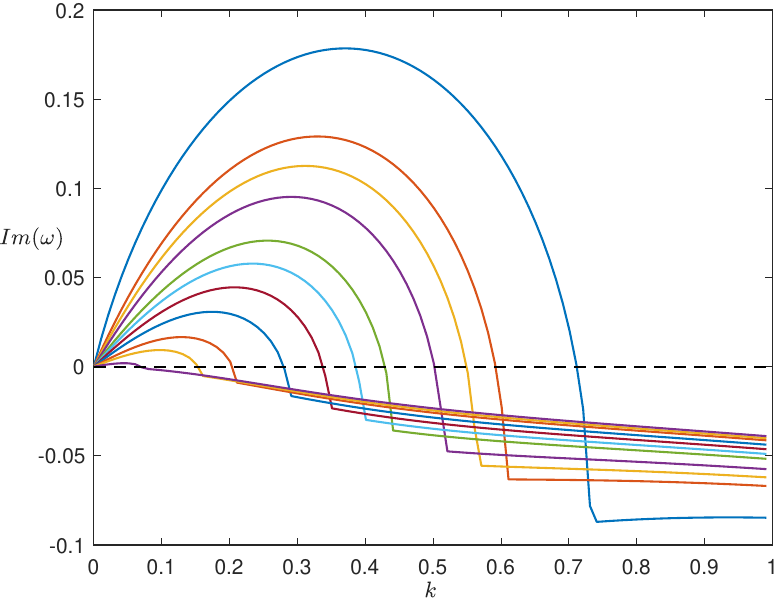}
	\includegraphics[width=0.4\columnwidth]{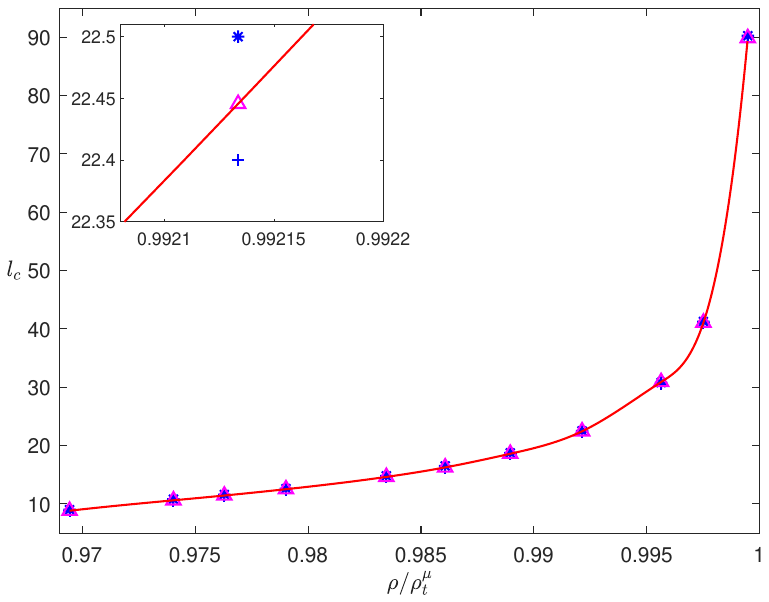}
	\caption{\textbf{The critical instability scale.}
		\textbf{Left: }$\mathrm{Im}(\omega)$ as a function of $k$ for the most unstable mode. The initial states of these colored lines are superfluid solutions in the range $\langle\mathcal{O}\rangle_t^{\rho}<\langle\mathcal{O}\rangle< \langle\mathcal{O}\rangle_t^{\mu}$. 
		\textbf{Right: }Numerical test of the critical scales. 
		Magenta triangles represent the critical scale extracted from the dispersion relations of the unstable modes presented in the left panel. The blue asterisk shows the length scale at and above which the initial solution runs away before $t_{max}$, while the blue cross show the length scale at and below which the initial solution does not evolve in time up to $t_{max}$.}\label{criticality}
\end{figure}

In order to do so, we select $11$ solutions in the range of $\langle\mathcal{O}\rangle_t^{\rho}<\langle\mathcal{O}\rangle< \langle\mathcal{O}\rangle_t^{\mu}$ as initial states and repeat this second set of numerical experiments. For each initial solution, we set the size of the box along the $x$ direction to different values, and see whether the infinitesimally deformed initial solution stays stable or run away to another final state in a long time $t_{max}=150000$. The results are shown in Figure~\ref{criticality}.

In the left panel of Figure~\ref{criticality}, we show the dependence of the imaginary part of the unstable mode on the wave vector $k$ for the $11$ solutions. The intersection of the curves with the horizontal axes indicates the critical value $k_c$, which is related to the critical length scale $l_c=2\pi/k_c$ of the system, beyond which the solution becomes unstable. Our results show that the solution which is closer to the turning point in the grand canonical ensemble has a larger value of the critical wave vector $k_c$, and therefore a smaller value of the critical length scale $l_c$. At exactly the turning point in the canonical ensemble, the critical wave-vector diverges and the finite size system becomes stable.

In the right panel of Figure~\ref{criticality}, we show the results of the dynamical test. The magenta triangles indicate the length scale calculated from the critical wave vector of the unstable mode from QNMs shown in the left panel of Figure~\ref{criticality}, and the red curve is an interpolation of the numerical data. The blue asterisks show the length scale at and above which the initial solution runs away before $t_{max}$, while the blue cross show the length scale at and below which the initial solution does not evolve in time in the range $0<t<t_{max}$. For practical purposes, we set the time scale $t_{max}=150000$. Near the critical length, the growth of the unstable modes takes an extremely long time. We can see that both the asterisk and cross are very close to the magenta triangle (inset in the right panel of Figure~\ref{criticality}), showing a perfect match between the results from quasinormal modes (QNMs) and this set of dynamical tests.

\begin{figure}
	\includegraphics[width=0.4\columnwidth]{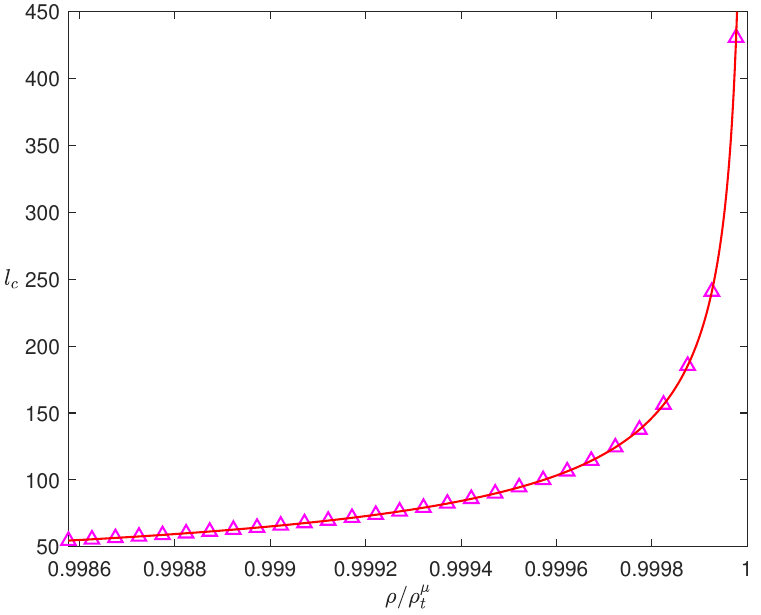}
        \caption{The critical scale $l_c$ near the turning point $\rho_t^\mu$ in the grand canonical ensemble. The magenta triangle represents the critical scale calculated from QNMs, while the red curve is the fitting result performed using Eq.\eqref{fitting_eq}.}\label{lc_fitting}
\end{figure}

When the solution approaches the turning point in the grand canonical ensemble, $\rho\rightarrow \rho_{t}^{\mu}$, the critical value of the wave vector becomes zero while the critical length scale diverges. It is interesting to fit this divergent behavior with a power-law function
\begin{align}
l_{c}(\rho)=R(1-\rho/\rho_{t}^{\mu})^{-\eta}.\label{fitting_eq}
\end{align}
In order to get better results, we add additional $29$ data points by calculating the QNMs in the region close to the turning point $\rho_{t}^{\mu}$, and present these data points (the magenta triangles) as well as the fitting curve (the red line) in Figure~\ref{lc_fitting}.

Fitting the data points with the three free parameters $R$, $\rho_{t}^{\mu}$ and $\eta$, as in Eq.\eqref{fitting_eq}, we obtain
\begin{align}
R\approx2.0317, ~~\rho_{t}^{\mu}\approx3.074, ~~\eta\approx0.5022.
\end{align}
Interestingly, $\eta$ is approximately $1/2$, which is the typical value of the critical exponent in mean field theory.

\subsubsection{Inhomogeneous final states with phase separation}\label{subsec6}
In the previous numerical experiments, we have  confirmed the non uniform instability as well as the existence of a critical length scale which can be predicted by the quasinormal mode dispersion. Another interesting question left to be investigated is whether the final state of the time dependent evolution triggered by the non uniform instability is also non uniform.

In principle, in order to answer this question, we would have to follow the full dynamical process up to the final state and then check whether the latter is homogeneous or not. However, because of charge conservation, it is possible to use the charge density of the uniform initial state to locate the position of the final state. With the full information of the QNM spectrum, we then know whether the possible uniform final state is stable. In case all the uniform solutions with the value of charge density equal to the initial one are unstable, then the real final state must necessarily be inhomogeneous. We will show such an example of this sort in the next subsection.

However, if we find that the final states are linearly stable, we are not able to conclude that the final state is homogeneous. Indeed, the final state might be a meta stable non-uniform phase which is linearly stable but nonlinearly unstable. In this case we have to run the full time dependent evolution to reach the final state and confirm this explicitly. Fortunately, within the holographic setup, we are able to run the full non-equilibrium evolution and analyze the nature of the final state. This is what the next subsection will be about. We anticipate that, only in some cases, the initial solutions with inhomogeneous instability end up into a uniform final state. In fact, we also find examples where the system reaches a non uniform final state at the end of the dynamical evolution.

\begin{figure}
	\includegraphics[width=0.23\columnwidth]{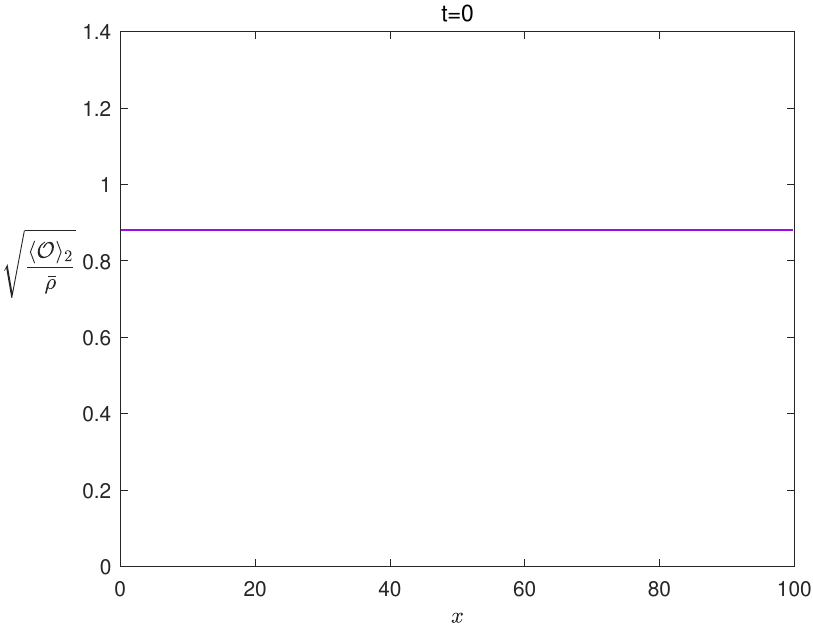}
	\includegraphics[width=0.24\columnwidth]{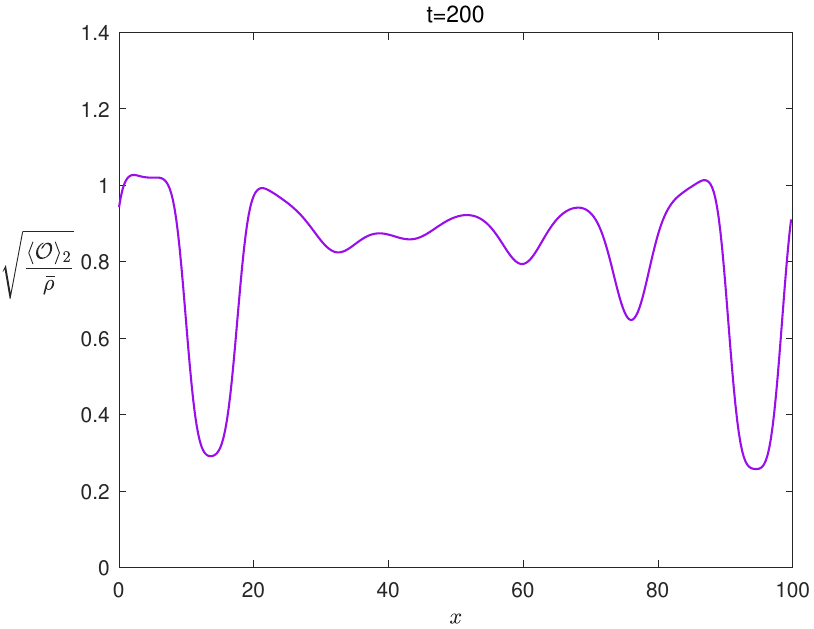}
	\includegraphics[width=0.24\columnwidth]{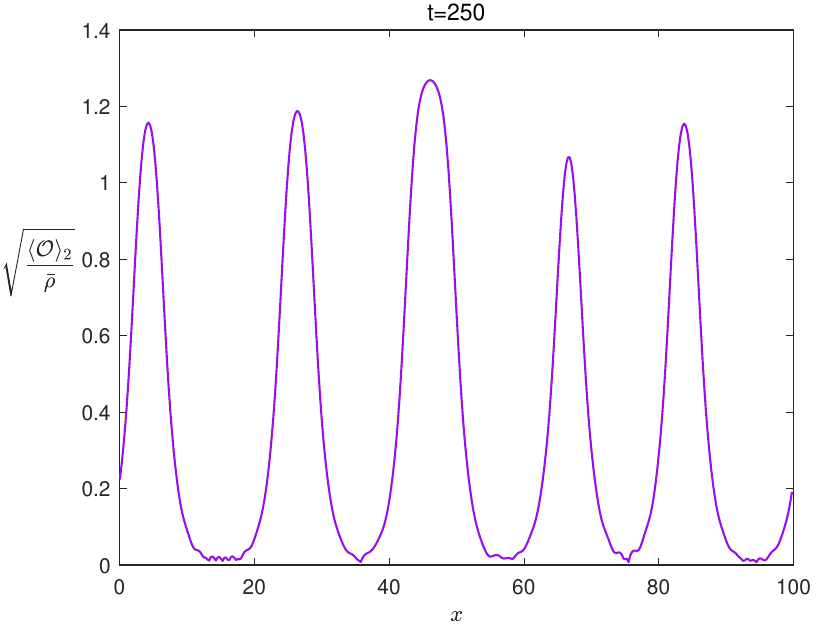}
	\includegraphics[width=0.24\columnwidth]{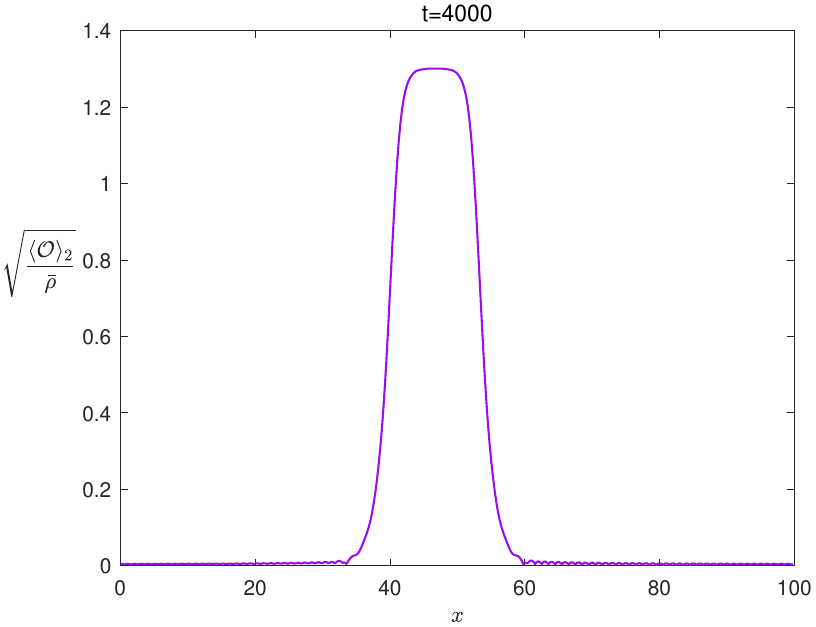}

 \vspace{0.2cm}
	
	\quad \includegraphics[width=0.23\columnwidth]{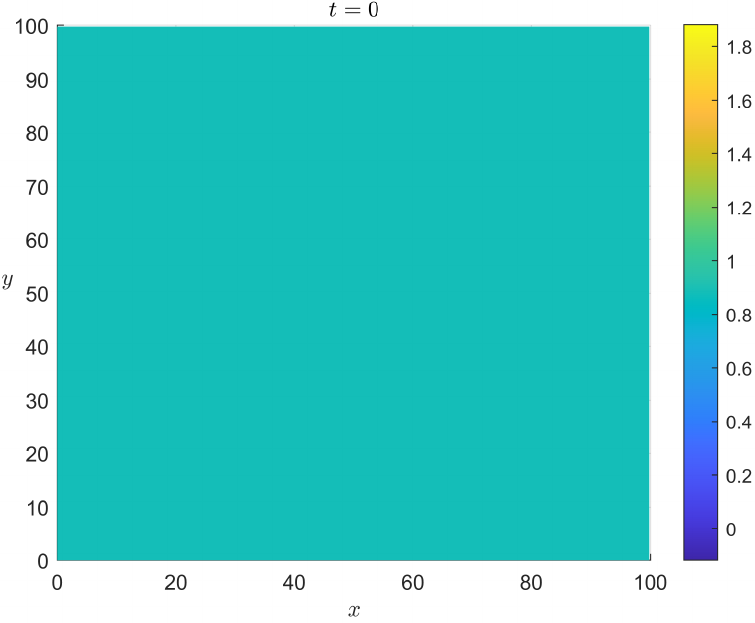}
	\includegraphics[width=0.23\columnwidth]{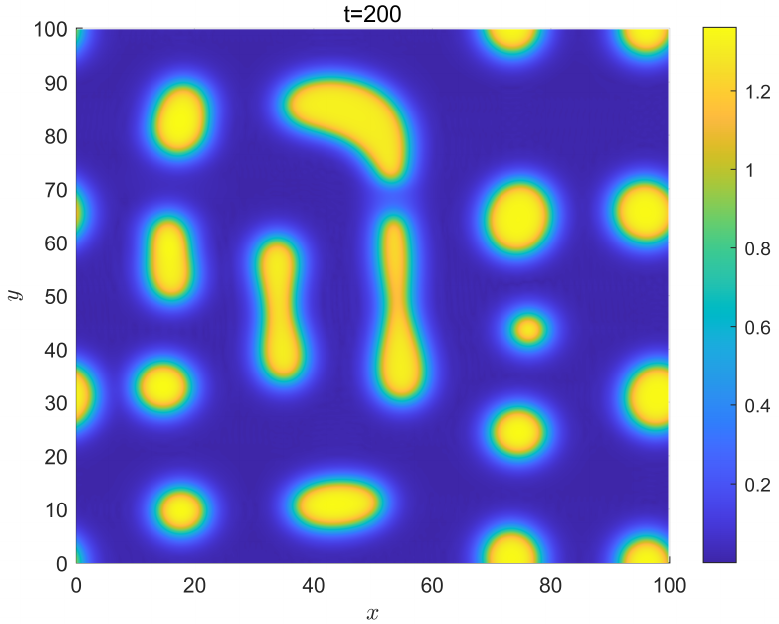}
	\includegraphics[width=0.23\columnwidth]{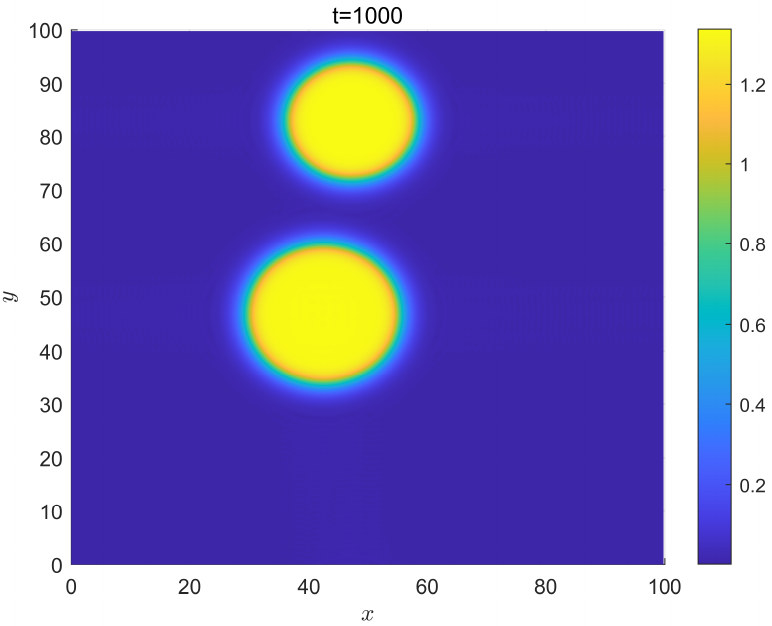}
	\includegraphics[width=0.23\columnwidth]{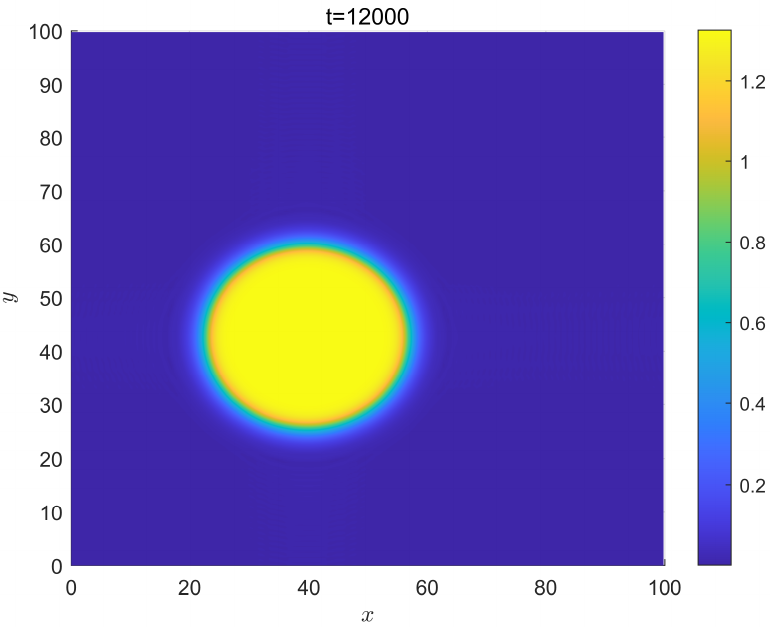}
	\caption{Local value of condensate at different time slices during the real-time evolution from a homogeneous initial state in one (\textbf{top panel}) and two (\textbf{bottom panel}) spatial dimensions with system size $l_x=l_y=100$. The initial state corresponds to the violet point in Figure \ref{1st_dynamic_k0} with charge density $\rho=3.00$.}\label{dynamic_process}
\end{figure}

We start by considering as initial state the solution marked by the purple point in Fig. \ref{1st_dynamic_k0} and set the box size to $l_x=l_y=100$. We then add an infinitesimal random perturbation to the initial state and let it evolve. For completeness, we have considered both 1D and 2D systems. The behavior of the condensate as a function of time during the full nonlinear time evolution is shown in Figure~\ref{dynamic_process} for the two cases. There, we clearly observed that during the time evolution bubbles are created. They then decrease in number and grow in size until the final state is reached. The final state contains a large symmetric bubble whose interior region is in a superfluid state with a very large value of the condensate. This implies that the final state is not homogeneous but displays evident phase separation between two regions (indicated in blue and yellow colors): one, inside the bubble, in the superfluid phase; the other, outside the bubble, in the normal phase with vanishing condensate.

From the real time evolution of the 1D and 2D systems, we see that these phase separation process, known as spinodal decomposition, is naturally triggered by the inhomogeneous linear instability appearing in the dispersion relation of the superfluid second sound mode. Interestingly, the whole dynamical evolution can be divided into four stages. In the first stage (or the early stage), the inhomogeneous unstable perturbations grow up exponentially as described by the linear analysis. In the second stage, the now large perturbations deform the homogeneous initial state and induce the formation of initial bubbles and domain wall structures. We see that the number of initial bubbles are determined by the scale of the wavelength of the most unstable mode. The allowed discrete values for the wave-vector $k$ in this finite system with $l_x=100$ are
\begin{equation}
k_{n}=2n\pi/l_x~,~n=0,1,2,...~.
\end{equation}
The maximum peak of the purple curve in the central panel of Figure ~\ref{1st_dynamic_k0} is located at $k_{m}=0.3119$ and indicates the most unstable wave-vector. We see that $k_m\approx k_5=5 k_1$ and therefore the most unstable mode is the one with wave-vector equal to $k_5$. The characteristic scale of the initial inhomogeneous state right after the onset of the instability is then given by $2\pi/k_5$, which is $1/5$ of the length scale of the whole system. Therefore, in the second stage we see approximately $5$ initial bubbles in the 1D system and $5\times5=25$ in the 2D one. This agrees very well with what observed in Figure \ref{dynamic_process}. In the third stage (or middle stage), the initial bubbles undergo a very diversified evolution. Some small bubbles shrink and disappear, while some big ones expand. Bubbles can also merge with each other creating a larger one. However, in most cases, when the chemical potential become balanced at each point, there is only one big bubble left in the final state. When only one big bubble is left, the system enters into the fourth stage (the final stage), in which the evolution of the big bubble becomes very slow. We notice that, occasionally, there might be two or more bubbles with the same radius left in the final state and when the radius of all bubbles become equal, the system goes into the fourth stage. The four stages just described are consistent with what suggested previously in Ref.~\cite{Attems:2019yqn}.

\subsection{Quench experiments}\label{subsec7}
We have already successfully realized phase separation from spinodal decomposition and analyzed its evolution in real time. However, in the previous ``experiments'', we have always started from unstable initial states. The condensate curve in the canonical ensemble (see left panel in Fig.\ref{1st_dynamic_k0}) reveals that it is possible to quench the system from a stable initial state towards an unstable state in the spinodal region. This might reflect a more realistic scenario which can be possibly compared with experimental setups.

\begin{figure}
	\includegraphics[width=0.35\columnwidth]{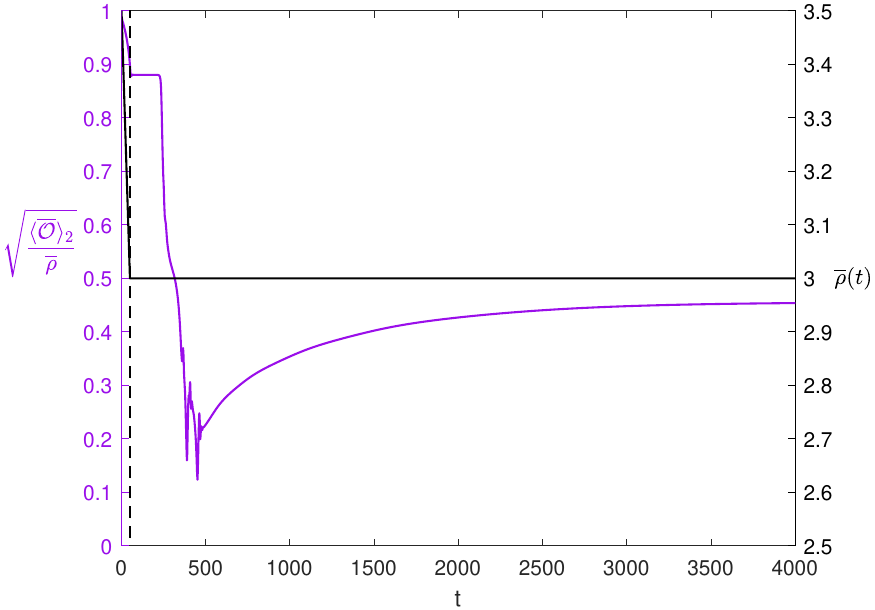}\qquad 
	\includegraphics[width=0.4\columnwidth]{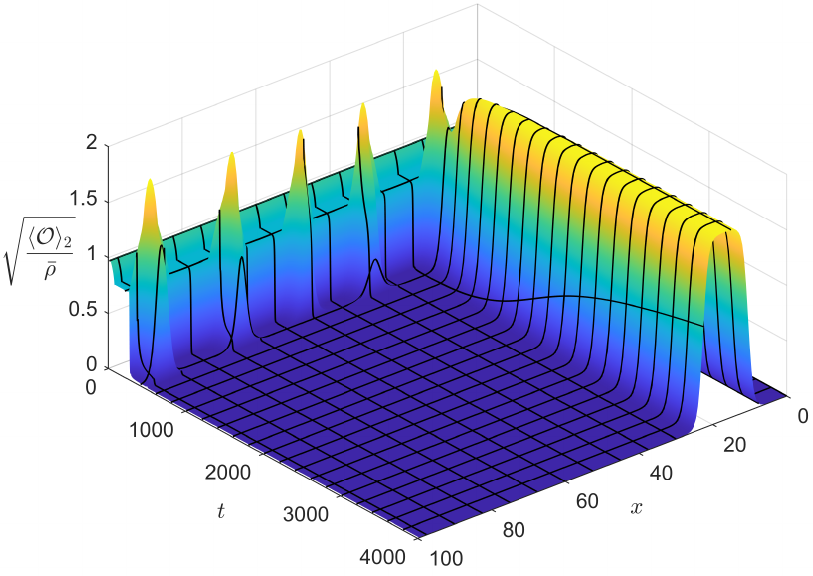}
	\caption{Time dependent quantities for a real time process with the average charge density quenched from $\rho_i=3.5005$ in the initial stable state to $\rho_f=3.00$, which is a value for the superfluid solution located in the spinodal region with $\langle\mathcal{O}\rangle_t^{\rho}<\langle\mathcal{O}\rangle< \langle\mathcal{O}\rangle_t^{\mu}$. The spatial size of the boundary system is $l_x=100$.
	\textbf{Left: }The solid violet curve is the average value of condensate as a function of time $t$ while the solid black curve is the average charge density $\bar{\rho}$ as a function of $t$, with dashed black vertical line denoting the end time of quenching the charge density. 
    \textbf{Right: }Local condensate as a function of time $t$ and space $x$.}\label{1DquenchA}
\end{figure}

\begin{figure}
	\includegraphics[width=0.23\columnwidth]{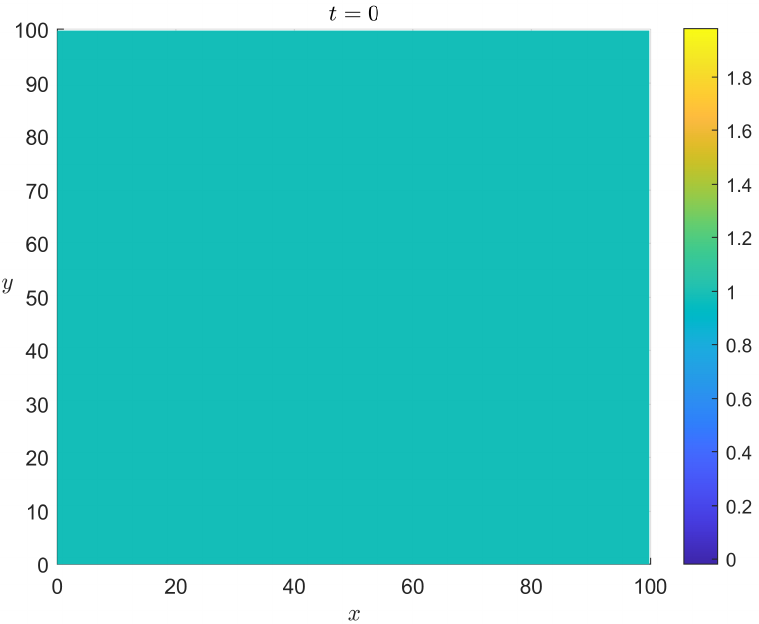}
	\includegraphics[width=0.23\columnwidth]{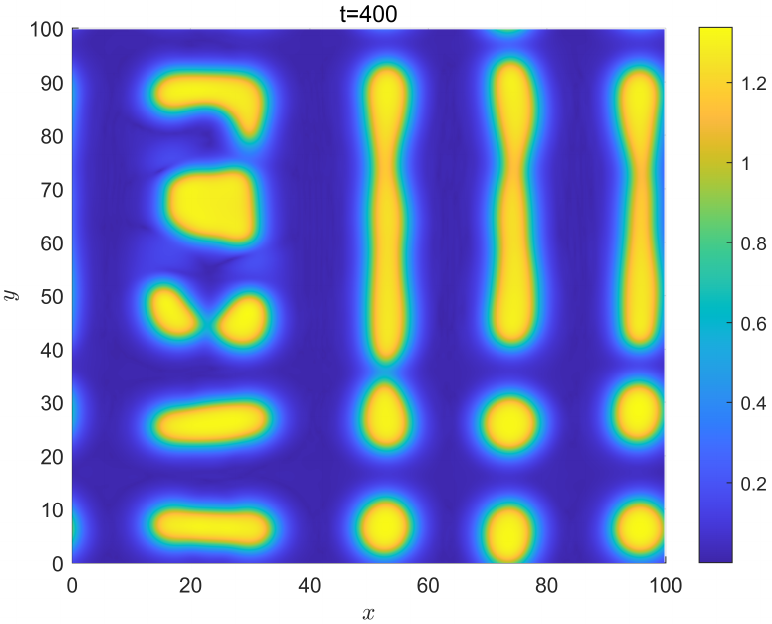}
	\includegraphics[width=0.23\columnwidth]{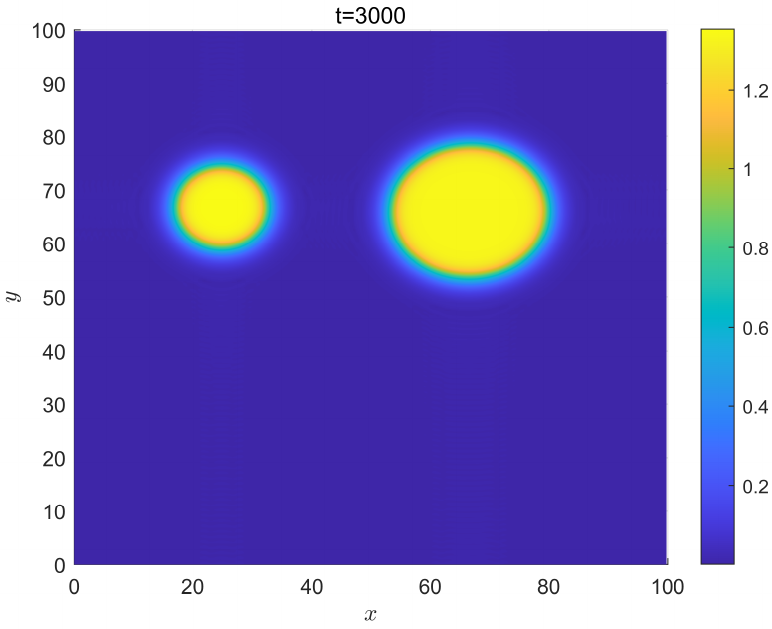}
	\includegraphics[width=0.23\columnwidth]{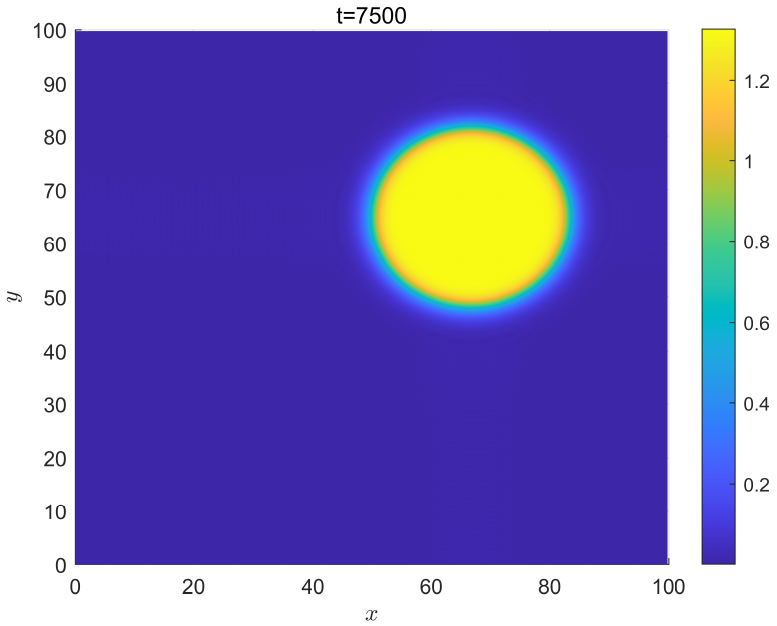}
\caption{Local condensate at four different time slices of a real time process with the average charge density quenched from $\rho_i=3.5005$ of the initial stable state to $\rho_f=3.00$ which is a value for the superfluid solution located in the spinodal region with $\langle\mathcal{O}\rangle_t^{\rho}<\langle\mathcal{O}\rangle< \langle\mathcal{O}\rangle_t^{\mu}$. In this quench process the two parameters for nonlinear terms are set to $\lambda=-2$, $\tau=0.8$.}\label{2DquenchA}
\end{figure}

Therefore, in this section, we perform additional numerical simulations to show the time evolution in these quench ``experiments''. In the first case with $\lambda=-2$, $\tau=0.8$, we quench the charge density of the system uniformly from a stable uniform initial state with $\rho_i=3.5005$ to the charge density value of the solution marked by the violet point in Figure \ref{1st_dynamic_k0}. The results of the $1$-dimensional numerical simulation are shown in Figure~\ref{1DquenchA}, while those of the $2$-dimensional numerical simulation are presented in Figure \ref{2DquenchA}. 

In Figure~\ref{1DquenchA}, we plot the time dependent value of the average charge density (solid black line) and average condensate (solid purple line) in the left panel, while we plot the time dependent value of local condensate in the right panel. We can see that in this quench experiment, the quench speed is fast, and the system first goes through the boundary of stability uniformly. Quickly after the end of the quench, the system first stays nearly uniform for a while, and then the spinodal decomposition appears. The domain walls emerge in the system due to the uneven disturbance. They then either disappear or grow until reaching a final equilibrium state.

In Figure \ref{2DquenchA}, we plot the 2D snapshot of the local condensate at four typical instants, where the creation and evolution of bubbles are evident. The quench speed is the same the 1D and 2D cases and the quenches are fast. Therefore, the qualitative behavior in this case is almost the same as the dynamical process from the unstable initial state in Figure~\ref{dynamic_process}. Similar to the results presented in the previous section,  the evolution can be also divided into four stages. In summary, the spinodal decomposition induced by a fast quench is qualitatively the same as that induced by considering an initial unstable state.

\color{black}

From the QNM spectrum of the static uniform states we can conclude that the final uniform normal solution is still stable under homogeneous and inhomogeneous perturbations. This solution also corresponds to the lowest value of free energy at this value of charge density. Nevertheless, we still cannot exclude the existence of a non-uniform meta-stable final state with lower value of the free energy. Therefore, in order to confirm the nature of final state, the full time evolution is necessary.

To compare with the previous case, we now consider a different system with $\lambda=-0.8$ and $\tau=0.28$, in which we are able to confirm the non-uniform final state without running the full dynamical evolution. In this case, the uniform solutions with the same value of charge density are all unstable under linear perturbations. In the rest of this section, we present the condensate curves for this case and introduce the logic that leads to the confirmation a non-uniform final state. The previous analysis for the critical length is also valid here. Nevertheless, since the results are analogous, to avoid clutter, we only give results for the 1-dimensional and 2-dimensional quench ``experiments'' and focus on the non-uniform final state.

\begin{figure}
	\includegraphics[width=0.28\columnwidth]{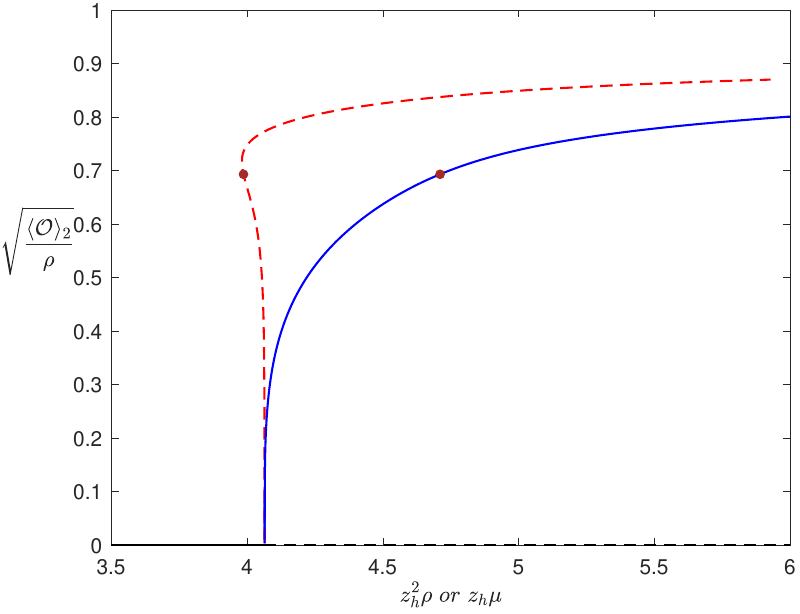}
	\includegraphics[width=0.30\columnwidth]{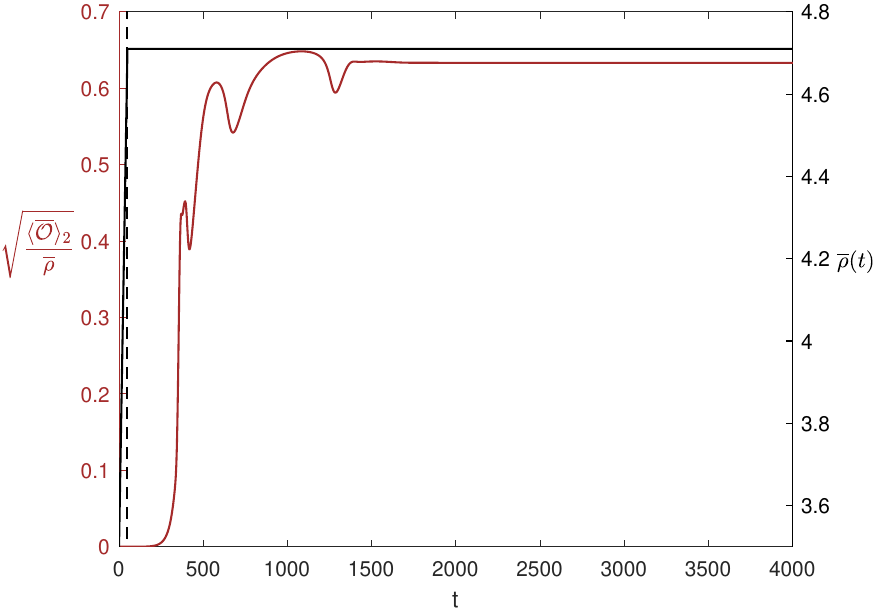}
	\includegraphics[width=0.35\columnwidth]{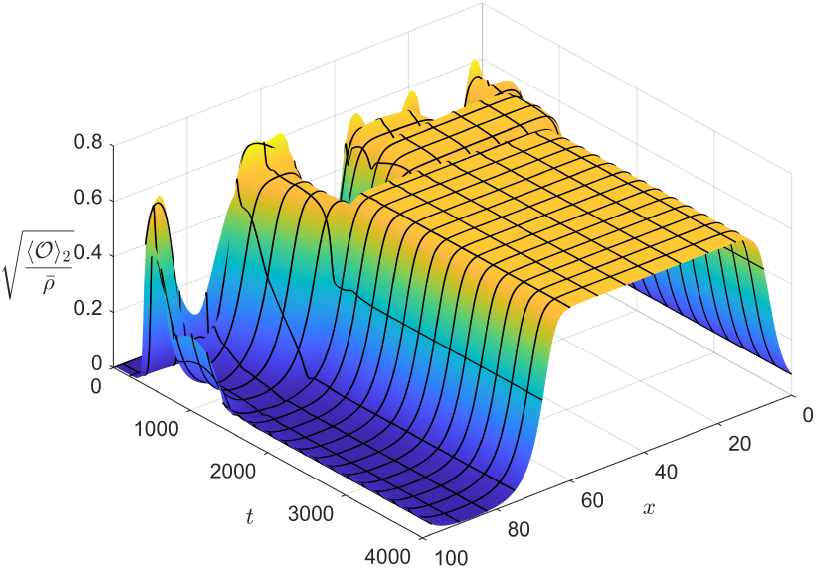}
	\caption{The condensate curves for uniform static superfluid solutions for $\lambda=-0.8$ and $\tau=0.28$ in two different ensembles and time dependent condensate in a one dimensional quench process from the initial stable normal state with $\rho_i=3.50$ to $\rho_f=4.71$ which is a value for the superfluid solution in the spinodal region. The size of the system is set to $l_x$=100.
	\textbf{Left: }The condensate as a function of the charge density $\rho$ in the canonical ensemble denoted by the solid blue line and the condensate as a function of chemical potential $\mu$ in the grand canonical ensemble denoted by the dashed red line. The brown point is the end point of the quench process with $\rho=4.71$.
	\textbf{Center: }The solid brown curve is the averaged value of the condensate as a function of time $t$. The solid black curve is the average charge density $\bar{\rho}$ as a function of $t$ and the dashed vertical black line indicate the end time of quenching $\rho$. 
	\textbf{Right: }Local condensate as a function of time $t$ and space $x$.}\label{quench2}
\end{figure}

We plot the condensate curves in the canonical and grand canonical ensembles in the left panel of Figure~\ref{quench2}, where we can see that the condensate curve in the canonical ensemble show a typical 2nd order phase transition, while the condensate curve in the grand canonical ensemble show a typical 1st order phase transition \cite{Zhang:2023uuq}. From the analysis of the QNMs performed in previous studies~\cite{Zhao:2022jvs}, we know that the solutions below the turning point in the grand canonical ensemble are stable under homogeneous perturbations, but unstable under inhomogeneous perturbations. Therefore, if we choose the final value of the averaged charge density equal to that of a solution in this region, and quench the system from a stable normal phase, we will certainly get a non-uniform final state exhibiting phase separation.

The main reason for this is that in the canonical ensemble, at this value of average charge density $\rho$, the normal phase is unstable, and the superfluid solution has lower value of the free energy. However, this superfluid solution is unstable when inhomogeneous perturbations are included and the total charge kept fixed. As a result, this homogeneous state with lowest value of free energy at this value of average charge density $\rho$ is also unstable. In summary, the final stable state must be some new non-uniform configuration.

To confirm this simple argument, we run a $1$-dimensional quench ``experiment'' and plot the average value of condensate versus time $t$ in the central panel of Figure~\ref{quench2}. At the same time, we provide a 3D figure to show the evolution of the local value of condensate in the right panel of Figure~\ref{quench2}. As expected, we observe that the final state is not homogeneous as it presents a large one-dimensional bubble. 

\begin{figure}
	\includegraphics[width=0.23\columnwidth]{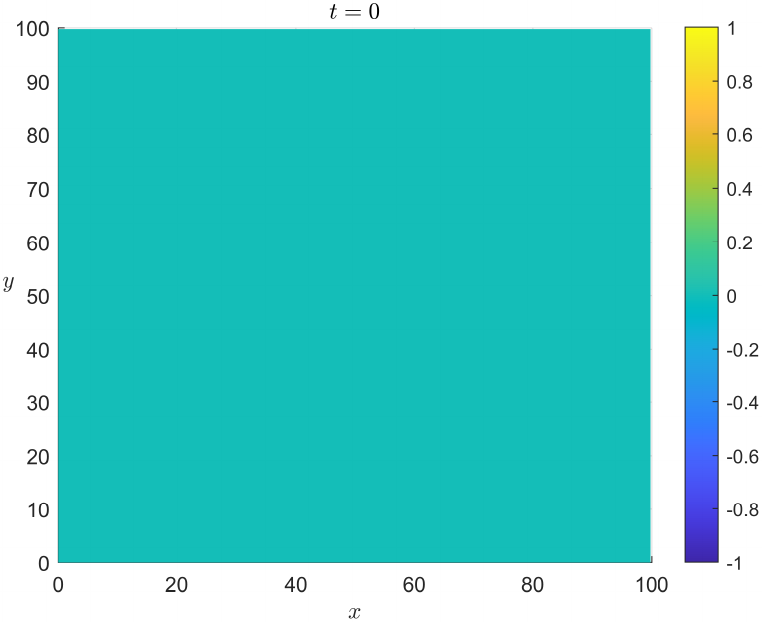}
	\includegraphics[width=0.23\columnwidth]{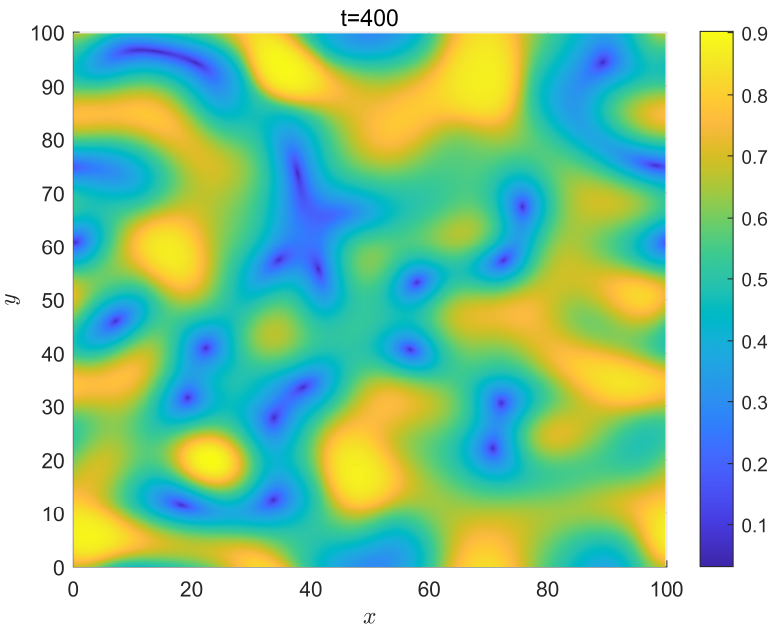}
	\includegraphics[width=0.23\columnwidth]{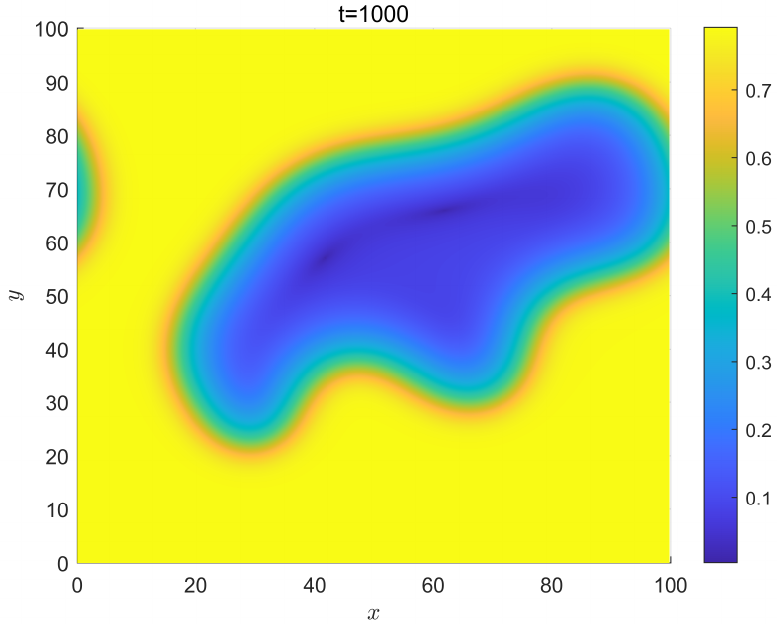}
	\includegraphics[width=0.23\columnwidth]{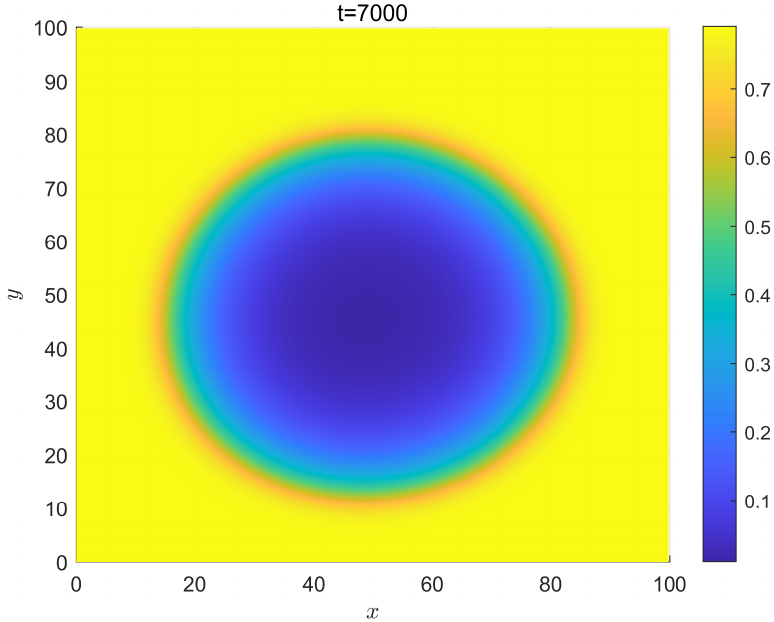}
\caption{The local condensate at four time slices in the two dimensional quench experiment from the initial stable normal state with $\rho_i=3.50$ to $\rho_f=4.71$ which is a value for the superfluid solution in the spinodal region for the case with $\lambda=-0.8$ and $\tau=0.28$.}\label{2DquenchB}
\end{figure}

Finally, we run a $2$-dimensional version of the same quench process and show the evolution of local condensate in four snapshots at different time in Figure~\ref{2DquenchB}. It is clear that the system runs into a non-uniform final state with a big bubble as well. Notice that the final bubble is a bubble of normal state inside a superfluid region, similar to a superfluid vortex, and different from the previous cases, \textit{e.g.}, the last panel in Figure \ref{2DquenchA}.

\begin{figure}
	\includegraphics[width=0.23\columnwidth]{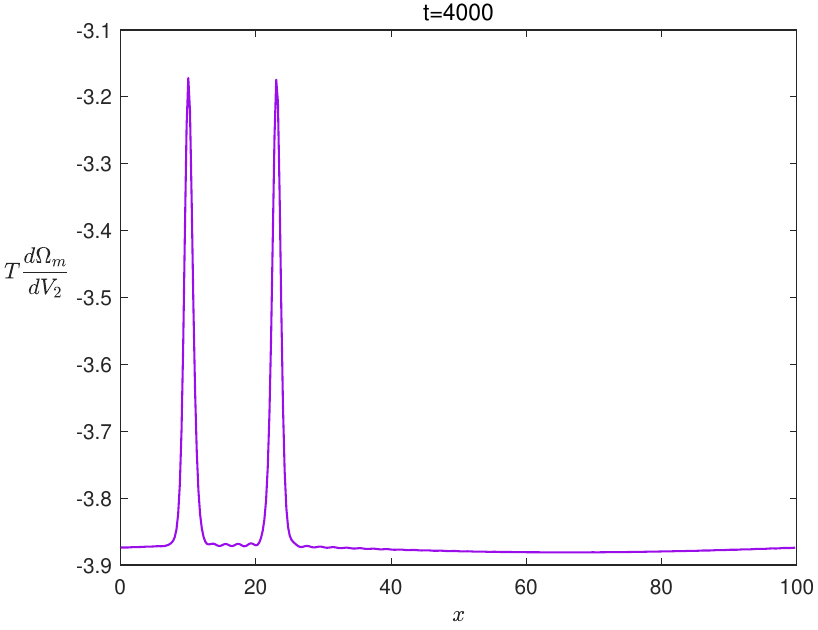}
	\includegraphics[width=0.23\columnwidth]{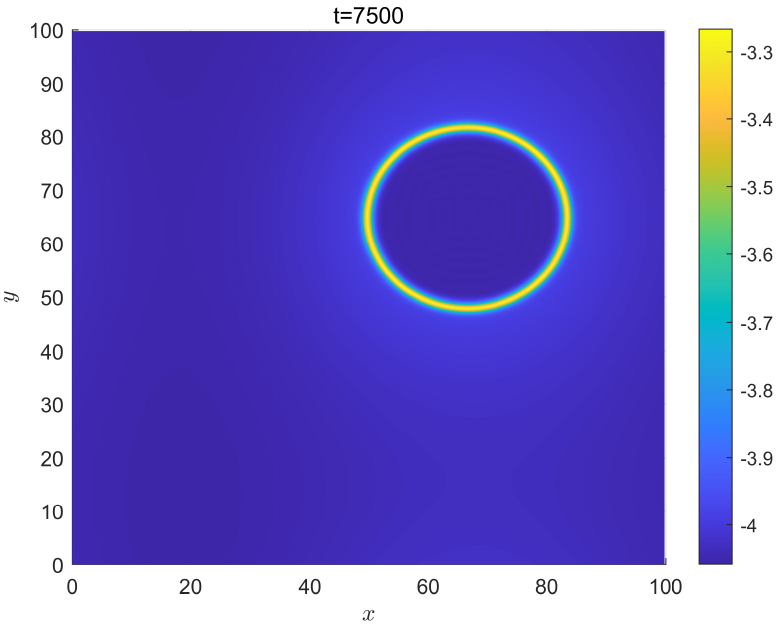}
	\includegraphics[width=0.23\columnwidth]{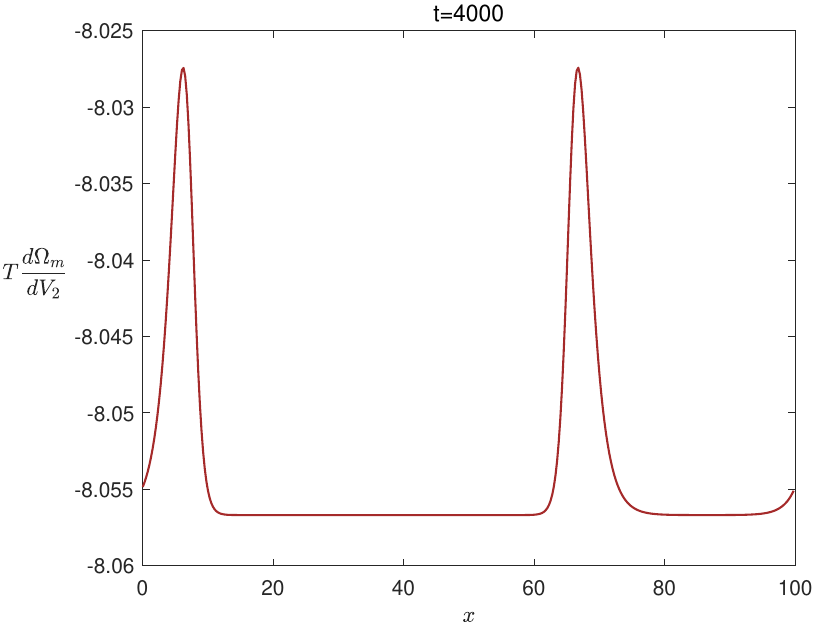}
	\includegraphics[width=0.23\columnwidth]{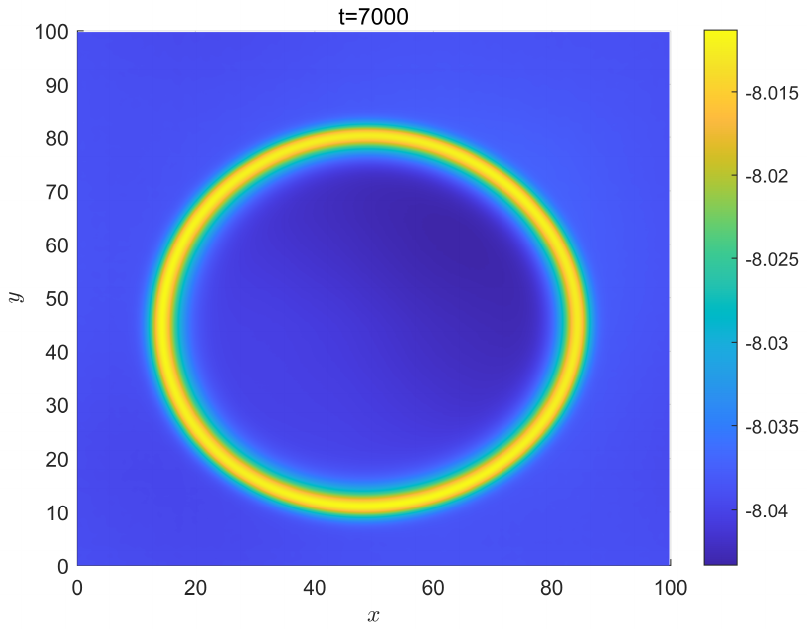}
\caption{\textbf{The grand potential density in the inhomogeneous finals states.} The grand potential density in the final states for the 1-dimensional and 2-dimensional quench experiments with ($\lambda=-2$, $\tau=0.8$) and ($\lambda=-0.8$, $\tau=0.28$).}\label{free_energy}
\end{figure}

The spatial distribution of the grand potential provides valuable insights, enabling us to conduct a more concrete analysis of this dynamical behavior. First, we show the local value of the grand potential density for the 1D and 2D quench experiments in Figure~\ref{free_energy}. The detailed formula for the grand potential density in in-going Eddington coordinate is presented in the Appendix, Eq.\eqref{A4}.\footnote{See also, \textit{e.g.} \cite{Tian:2019fax} for a detailed discussion of the free energy and grand potential in the context of holography.}

In Figure~\ref{free_energy}, the first and second panels depict the grand potential density configurations of the 1-dimensional and 2-dimensional final states for the case $\lambda=-2$ and $\tau=0.8$, respectively. The third and fourth panels show the relative results for the case $\lambda=-0.8$ and $\tau=0.28$. These configurations clearly show the grand potential density barrier along the bubble walls. The height of the wall is larger in the first case and much lower in the second case, which is important in explaining the difference in the phase separation dynamics. In the first case, small bubbles emerge quickly in the second stage of time evolution. While in the second case, small bubbles do not show up clearly, and some connected narrow region emerge instead. The reason behind this difference is that in the first case the energy barrier of the bubble wall is large, and the decrease of the length of bubble wall is efficient in minimizing the total grand potential. However, in the second case, the height of the energy barrier on the bubble wall is much lower, and the width of the bubble wall is larger. In this case the phase separation process is not likely to produce many small bubbles. The detailed relation between the height of the bubble wall and the creation and evolution of small bubble is worth to be studied more carefully in the future.

Another important issue is that during the 1-dimensional evolution, the chemical potential inside and outside the bubble has to be balanced. In the 2-dimensional case, because of the different dependence of the surface and volume terms on the radius of the bubble, the grand potential density inside a stable bubble should be lower than the grand potential density outside \cite{Li:2020ayr}. This feature is also evident in the four plots in Figure~\ref{free_energy}.

\begin{figure}
\quad\quad
        \includegraphics[width=0.4\columnwidth]{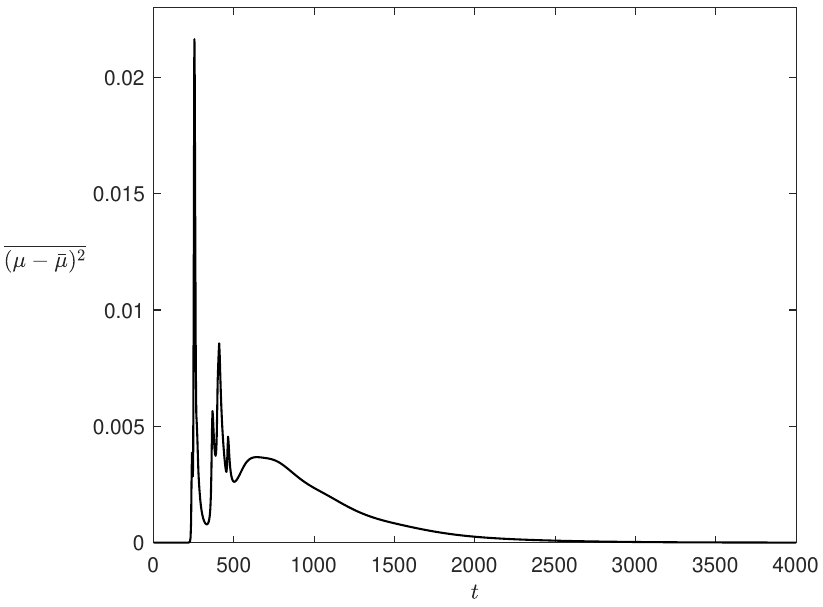} \quad
	\includegraphics[width=0.4\columnwidth]{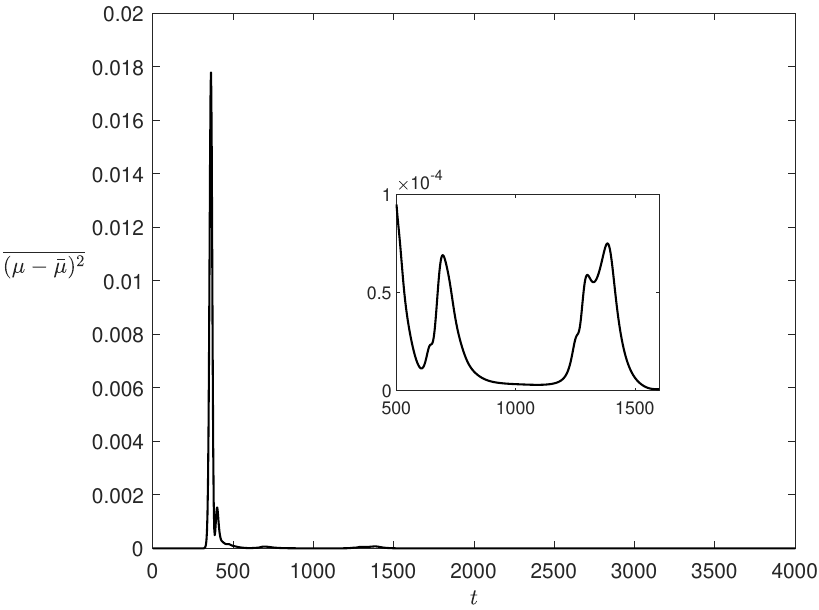}

 \vspace{0.4cm}
 
        \quad \quad \includegraphics[width=0.4\columnwidth]{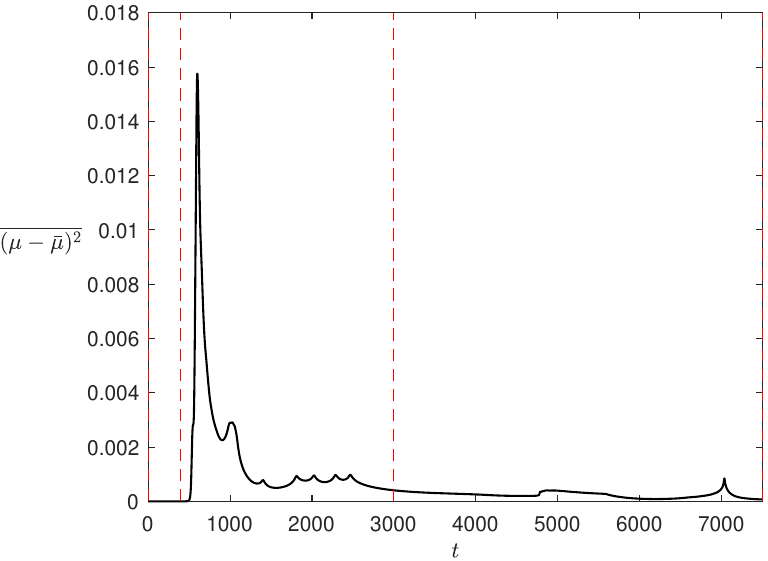} \quad 
        \includegraphics[width=0.4\columnwidth]{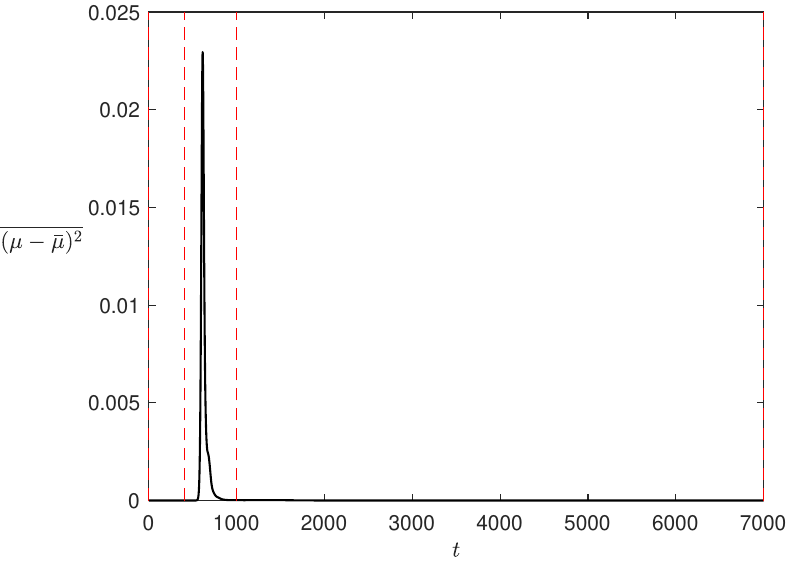}
\caption{\textbf{Dynamical heterogeneity during the out-of-equilibrium evolution.} The averaged value of the variance of the local value of chemical potential $\overline{(\mu-\bar{\mu})^2}$. The top two panels denote the evolution of the variance of $\mu$ in one-dimensional quenching processes while the bottom two panels denote the relation in two-dimensional quenching processes. The two left panels are for the case with  ($\lambda=-2$, $\tau=0.8$) and the right two panels are for the case with ($\lambda=-0.8$, $\tau=0.28$). We find that the peaks of the curves correspond to the generation or disappearance of domain walls and bubbles. In the bottom plots, we use four dashed vertical red lines to denote the times at which we plotted the snapshots of local condensate in Figure \ref{2DquenchA} and Figure \ref{2DquenchB}.}\label{meanvarianceA}
\end{figure}

For convenience, in the numerical simulations, we fix the gauge by setting $A_{t}|_{z=0}$ to a constant independent of $t$ and $\vec{x}$, instead of using the usual gauge choice $A_{t}|_{z=1}=0$ as in the static case. For any final equilibrium state, including the non-uniform states, the chemical potential, defined as
\begin{align}
	\mu(t,\vec{x})=A_{t}|_{z=0}(t,\vec{x})-A_{t}|_{z=1}(t,\vec{x})~\label{eq_mu},
\end{align}
should be balanced everywhere -- the so-called ``chemical balance condition'' -- as proved in \cite{Li:2020ayr}. However, during the non-equilibrium evolution, the chemical potential varies in space. Therefore, the variance of the chemical potential defined as
\begin{align}
	\delta\mu=\overline{(\mu-\bar\mu)^2}
\end{align}
is a good measure of the degree of inhomogeneity during the time evolution -- dynamical heterogeneity.
We plot the spatial variance of chemical potential $\delta \mu$ for the 1-dimensional and 2-dimensional quench experiments in Figure~\ref{meanvarianceA}. The left two panels are for the 1-dimensional (top panel) and 2-dimensional (bottom panel) quench experiments for the first case with ($\lambda=-2$, $\tau=0.8$). The two panels on the right side are for the 1-dimensional (top panel) and 2-dimensional (bottom panel) quench experiments for the second case with ($\lambda=-0.8$, $\tau=0.28$). In the two panels for the 2-dimensional processes, we use four dashed vertical red lines to indicate the times corresponding to the four snapshots of local condensate presented above in Figure \ref{2DquenchA} and Figure \ref{2DquenchB}. We see that the variance of chemical potential $\delta\mu$ reaches an initial maximum, which roughly corresponds to the exponential growth of the unstable modes. After the initial peak, there are some following smaller peaks, which corresponds to the regime in which small bubbles merge and disappear. The sub peaks are sharper in the 1-dimensional cases than in the 2-dimensional cases, because in the 1-dimensional case, the domain wall region expands into a larger portion of the whole volume. Finally, the system gradually approaches the stable final state forming a big bubble, and the variance of chemical potential $\delta\mu$ decreases to $0$ at that point.

\section{Outlook}\label{sect:conclusion}
In this work, we have performed several numerical experiments on the full dynamical evolution triggered by the spinodal instability in first order phase transitions in a holographic superfluid model.

In the first part of this work, we focused on confirming the linear dynamical stability analysis (based on the QNM spectrum) in the spinodal region. This critical value of wave vector $k_c$ identifies a critical value for the size of the system, $l_c=2\pi/k_c$, below which the system is stable. This prediction has also been confirmed explicitly by our numerical ``experiments'' involving the fully nonlinear time evolution of the system.

In the second part, we presented the results of the full time dependent evolution in order to analyze in more detail the out-of-equilibrium dynamics, as well as the nature of the final state. We have considered both 1-dimensional and 2-dimensional quench ``experiments'' for two different configurations of the potential in the model. In the real time evolution from both the unstable initial state and the quench processes, the inhomogeneous instabilities trigger the spinodal decomposition. The following time evolution can be divided into four typical stages as suggested in Ref.~\cite{Attems:2019yqn}. The local value of grand potential for the final stable state is consistent with the general picture of bubbles. We also find that the variance of chemical potential $\delta\mu$ is a good indicator to describe the dynamical heterogeneity of the system during the real time evolution. There, a clear peak in the variance is observed at early times and is related to the sudden growth of the inhomogeneous instabilities. Following sub peaks are accompanied by the disappearance and merge of small bubbles.

Our study provides a concrete example of the dynamical evolution of a system under spinodal decomposition, which is triggered by the inhomogeneous linear instability in a region with negative susceptibility $\partial \rho/\partial \mu <0$, when the size of the system is larger than some critical length. The initial and final regimes of the time evolution perfectly agree with the results of the linear dynamical stability based on quasi-normal modes~\cite{Zhao:2022jvs}. Our study also shows that the two thermodynamic ensembles (canonical and grand-canonical) are both necessary in order to understand the dynamical processes behind the spinodal decomposition. Finally, we also find that quenching the extensive variable $\rho$ of the system is a perhaps more realistic way to reach the unstable spinodal region and study the full dynamics of the phase separation triggered by the spinodal decomposition in such a setup.

There are many open questions left for the future. In the quenching experiment from a normal state to the final superfluid phase, the coexistence of two different topological defects, the bubbles and the superfluid vortexes, is expected in a large system. The detailed relation and effective interaction between the two is not known and can be studied using holographic methods in the future. Also, a more detailed analysis of the evolution of the energy power spectrum during phase separation might disclose important information about gravitational wave production from cosmological first order phase transitions. Finally, it would also be interesting to study the phase separation with a conserved total magnetic flux, and investigate its influence on the bubbles and vortexes. We are planning to pursue some of these directions in the near future.

\section*{Acknowledgements}
ZYN would like to thank Rong-Gen Cai and Chuan-Yin Xia for useful discussions. ZYN would also like to thank the organizers of ``2023 International Workshop on Gravity and Cosmology'' in Lanzhou University for their hospitality. This work is partially supported by NSFC with Grant No.11965013, 11881240248, 11975235, 12035016 and 12375058. ZYN is partially supported by Yunnan High-level Talent Training Support Plan Young $\&$ Elite Talents Project (Grant No. YNWR-QNBJ-2018-181). MB acknowledges the support of the Shanghai Municipal Science and Technology Major Project (Grant No.2019SHZDZX01) and the sponsorship from the Yangyang Development Fund.

\appendix
\section{Numerical Scheme}\label{appendix1}
During the time dependent evolution, the value of $A_{t}$ at a given time slice is calculated from Eq.\eqref{equation5}, while the other field components are evaluated from previous time slice by the fourth-order Runge-Kutta(RK4) method with a time step $\Delta$t=0.05 with Equations (\ref{equation1}-\ref{equation4}). Filtering of the high momentum modes is implemented following the usual ``2/3’s rule'', with the uppermost one-third of Fourier modes being removed \cite{Chesler:2013lia}.

The restriction given by Eq.\eqref{equation2} is applied as a boundary condition at z$\rightarrow$0 for $A_{t}$, which simplifies to
\begin{align}
	&\partial_t\rho=-\partial_x(\partial_zA_x+\partial_x \phi)-\partial_y(\partial_zA_y+\partial_y \phi)~.\label{A1}
\end{align}
The conserved current is
\begin{align}
J^\mu=-\partial_z A_\mu-(\partial_\mu A_t-\partial_t A_\mu)~,
\end{align}
where $\mu=x, y$. Using the boundary conditions $A_\mu|_{z \rightarrow0}=0$, one can derive that Eq.\eqref{A1} concides with the conservation equation for the U(1) charge
\begin{align}
\partial_\mu J^\mu-\partial_t\rho=0
\end{align}
at the boundary.

In the radial z-direction, we use the Chebyshev spectral method with $21$ grid points. For the boundary spacial directions $x$ and $y$, we set periodic boundary conditions, and use the Fourier spectrum method with the spatial spacing $\Delta x=0.2$. For the dynamic evolutions considered in this work, we choose uniform solutions as the initial states, which satisfy the following constraints in Eddington coordinates
\begin{align}
	f\partial_z \partial_z \psi+f'\partial_z \psi+2i\phi \partial_z \psi-z\psi+i\psi\partial_z \phi-2\lambda \psi^2 \psi^*-3\tau z^2 \psi^3 \psi^{*2}=0,\label{A2}\\
	\partial_z \partial_z \phi+i\psi \partial_z \psi^*-i\psi^* \partial_z \psi=0,\label{A3}
\end{align}
where the prime denotes the derivative with respect to $z$.

It is worth noting that in the numerical simulation, even if we do not add small perturbations by hand, fluctuations below the precision  of floating point numbers still exist. Therefore, to better control the small perturbations, we add finite random perturbations at the level $10^{-12}$ on top of our initial solution as well.

The grand potential density on a time slice is calculated in the ingoing Eddington coordinates as follows:
\begin{equation}
\begin{aligned}
\frac{d\Omega_m}{d V_2}&=\frac{L^{2}}{T}\Big[-\frac{\mu \rho}{2}-\left(\frac{A_t \partial_z A_t}{2}\right)|_{z=z_h}+\int_{0}^{z_h}(\psi^* \partial_x \partial_x \psi+\psi^* \partial_y \partial_y \psi+\partial_x \psi \partial_x \psi^*+\partial_y \psi \partial_y \psi^*-\partial_z \psi^* \partial_t \psi-\partial_z \psi \partial_t \psi^* \\
&\quad-2 \psi^* \partial_t \partial_z \psi-A_x \partial_t \partial_z A_x-A_y \partial_t \partial_z A_y-\partial_z A_x \partial_t A_x-\partial_z A_y \partial_t A_y-\partial_y A_x \partial_x A_y+\partial_x A_t \partial_z A_x+\partial_y A_t \partial_z A_y \\
&\quad+i q \psi \psi^* \partial_z A_t-i q \psi \psi^* \partial_x A_x-i q \psi \psi^* \partial_y A_y-q^2 A_x^2 \psi \psi^*-q^2 A_y^2 \psi \psi^*+\frac{1}{2} (\partial_y A_x)^2+\frac{1}{2} (\partial_x A_y)^2 \\
&\quad+\frac{1}{2} A_x \partial_y \partial_y A_x+\frac{1}{2} A_y \partial_x \partial_x A_y-\frac{1}{2} i q A_x \psi \partial_x \psi^*-\frac{1}{2} i q A_y \psi \partial_y \psi^*-\frac{1}{2} A_x \partial_x \partial_y A_y \\
&\quad-\frac{1}{2} A_y \partial_x \partial_y A_x+\frac{1}{2} A_x \partial_z \partial_x A_t+\frac{1}{2} A_y \partial_z \partial_y A_t+\frac{1}{2} A_t \partial_z \partial_x A_x+\frac{1}{2} A_t \partial_z \partial_y A_y \\
&\quad+\frac{1}{2} i q A_t \psi \partial_z \psi^*-\frac{3}{2} i q A_x \psi^* \partial_x \psi-\frac{3}{2} i q A_y \psi^* \partial_y \psi+\frac{3}{2} i q A_t \psi^* \partial_z \psi \\
&\quad-\frac{\psi^* \partial_t \psi}{z}-\frac{\psi \partial_t \psi^*}{z}-\lambda \psi^2 \psi^{*2}-2z^2\tau \psi^3 \psi^{*3})dz\Big].\label{A4}
\end{aligned}
\end{equation}

\bibliographystyle{apsrev4-1}
\bibliography{reference}

\end{document}